\documentstyle[pre,aps,multicol,eqsecnum,tighten,epsf]{revtex}

\begin{document}
\draft

\title{Propagation and Structure of Planar Streamer Fronts}
\author{Ute Ebert and Wim van Saarloos}
\address{
Instituut--Lorentz, Universiteit Leiden, Postbus 9506, 2300 RA Leiden, 
the Netherlands,
       }
\author{ Christiane Caroli}
\address{
Universit\'e Paris VII, GPS Tour 23, 2 Place Jussieu, 75251 Paris Cedex 05,
France       }
\maketitle

\begin{abstract}
Streamers are a mode of dielectric breakdown of a gas in a 
strong electric field: A sharp nonlinear ionization wave 
propagates into a non-ionized gas, leaving a nonequilibrium 
plasma behind. The ionization avalanche in the tip of the wave 
is due to free electrons being accelerated in the strong field 
and ionizing the gas by impact. This chain reaction deeper in 
the wave is suppressed by the generated free charges screening 
the field. Simulations of streamers show two widely
separated spatial scales: the width of the charged layer where the 
electron density gradients and the ionization rate are very large
(${\cal O}(\mu$m)), and the width of the electrically 
screened, finger-shaped and ionized region (${\cal O}$(mm)). 
We thus recently have suggested to analyze first the properties 
of the charge-ionization-layer on the inner scale on which it is 
almost planar, and then to understand the streamer shape on the outer 
scale as the motion of an effective interface as is done in other 
examples of nonequilibrium pattern formation. 
The first step thus is the analysis of the inner dynamics of 
planar streamer fronts. For these, we resolve the long-standing
question about what determines the front speed, by applying
the modern insights of pattern formation to the
streamer equations used in the recent simulations. These include 
field-driven impact ionization, electron drift and diffusion and 
the Poisson equation for the electric field. 
First, in appropriately chosen dimensionless units 
only one parameter remains to characterize the gas, the 
dimensionless electron diffusion constant $D$; for typical gases 
under normal conditions $D\approx0.1$-0.3. 
Then we determine essentially all relevant properties of planar 
streamer fronts. Technically, we identify the propagation of 
streamer fronts as an example of {\em front propagation into unstable 
states}. In terms of the marginal stability scenario we then find: 
the front approached asymptotically starting from any sufficiently 
localized initial condition (the ``selected front''), 
is the steepest uniformly translating front solution, which is 
physical and stable. Negatively charged fronts are selected by 
linear marginal stability, which allows us to derive their velocity 
analytically. Positively charged fronts can only propagate due to 
electron diffusion against the electric field; as a result their 
behavior is singular in the limit of $D \to 0$. For $D \lesssim 1$, 
these fronts are selected by nonlinear marginal stability and we 
have to apply numerical methods for predicting the selected front 
velocity. For larger $D$, linear marginal stability applies and the 
velocity can be determined analytically. 
Numerical integrations of the temporal evolution of planar fronts out of 
localized initial conditions confirm all our analytical and numerical 
predictions for the selection. Finally, our general predictions for the
selected front velocity and for the degree of ionization of the plasma are 
in semi-quantitative agreement with recent numerical solutions of 
three-dimensional streamer propagation. This gives credence to our
suggestion, that the front analysis on the inner ($\mu$m) scale yields
the moving boundary conditions for a moving ``streamer interface'',
whose pattern formation is governed by the evolution of the fields
on the outer (mm) scale.
\end{abstract}

\pacs{47.54.+r, 52.80.Mg, 51.50.+v}

\begin{multicols}{2}

\section{Introduction}

Discharges are nonequilibrium ionization processes occurring in
initially non-ionized matter exposed to strong electric fields.
Depending on the spatio-temporal characteristics of the electric field
and on the ionization and charge transport properties of the medium,
discharges can assume many different modes of appearance. In
particular in gases under approximately normal conditions one
distinguishes phenomenologically between stationary modes like arc,
glow or dark discharges and transient phenomena like leaders, the 
initial stages of sparks, and streamers \cite{Gal,Gal2,Raizer,Lagar,vasilyak,footnote1}, which
occur e.g. in silent discharges \cite{silent}. The
latter nonstationary discharges often form the initial state of a
discharge that later on becomes stationary. We will focus here on an
essential element of many transient discharge phenomena, the initial
field-driven ionization wave.

The conceptually simplest problem of this kind has become known as the
streamer problem in a non-attaching gas. It treats the dynamics of the
free electrons and positive ions in a homogeneous gas at rest taking
the following mechanisms into account: {\em (i)} impact ionization,
the process in which a free electron accelerated in a strong local
field ionizes a neutral molecule, generating a new free electron and a
positive ion; {\em (ii)} drift and diffusion of charged particles, in
particular of the electrons whose mobility is much larger than that of
the ions; {\em (iii)} the coupling of the electric field to the
charges through the Poisson equation of electrostatics.

Recent numerical simulations \cite{DW,Vit} of a basic model
incorporating these physical ingredients for parameter values
appropriate for nitrogen under normal conditions reveal that a
streamer consists of a sharp nonlinear ionization front which
propagates into a non-ionized gas, leaving a weakly ionized
nonequilibrium plasma behind. The underlying mechanism is that in the
leading edge of the front the electrons are accelerated by the large
imposed electric field; this causes the build-up of an electron
avalanche due to impact ionization.  The generated free charges
eventually screen the field and thus suppress further ionization. It
is the nonlinear balance between these two nonequilibrium processes,
namely the ionization avalanche and the electric screening, which
determines the dynamics of the ionization front and the state of the
plasma behind it. In confined geometries, streamers usually have a
nontrivial fingerlike shape, as is illustrated by the snapshots in
Fig.~1 of streamer dynamics taken from the simulations of Vitello {\em
  et al.} \cite{Vit}. As the sharpness of the electron density
profiles in Fig.~1 illustrates, the ``passive body'' of the finger is
separated from the external non-ionized gas by a very narrow region
--- of width of order microns --- in which essentially all the action
is occurring. This width has to be compared to the size of the
filament, which is of order millimeters. It is in this narrow layer
that most of the ionization process is taking place. In this same
region, there is a nonzero charge density, and consequently, also a
very large electric field gradient.  These features indicate that
there are two different spatial scales in this process, an ``inner''
scale associated with the thickness of the zone where the ionization
takes place, and an ``outer'' one where the spatial variations are set
by the size of the finger and the external experimental geometry.  It
is precisely for these reasons that accurate simulations are extremely
demanding and that they were accomplished only recently by Dhali and
Williams \cite{DW} and by Vitello {\em et al.} \cite{Vit} (See also
\cite{wan}).

Such a separation of scales is strongly reminiscent of what occurs in
combustion fronts \cite{wil,buc}. A combustion front is a narrow
layer of thickness $\ell_{in}$ to which the combustion is essentially
confined, while outside of it, the temperature field varies on a much
longer scale $\ell_{out}$. Physically, such sharp combustion fronts
occur in the limit when the chemical reaction rates involved in the
combustion are very fast once a sufficient temperature is reached. It
has been shown that, on the basis of an asymptotic expansion to lowest
order in the small parameter $\varepsilon= \ell_{in} / \ell_{out}$
using matched asymptotic expansions \cite{vDyke,bender}, the
problem can be analyzed in terms of the propagation of an ``effective
interface''. More specifically, one first solves the so-called inner
problem of a locally {\em almost planar} reaction zone. This permits to
relate the temperature and chemical composition fields on both sides
of the front (at distances $L$ such that $\ell_{in} \ll L \ll
\ell_{out}$) and to determine the local front velocity as a function
of local curvature and fields. On the scale of the remaining outer
problem, these relations then play the role of boundary conditions 
and of a kinetic equation for the effective moving interface of zero 
thickness. Besides in combustion, the technique of asymptotic
matching to obtain an effective interface description has also
been applied to chemical waves \cite{Meron}, thermal plumes
\cite{ZTB} and to phase field models of solidification \cite{pf1,pf2}.

In spite of some important differences between combustion and streamer
fronts as discussed in the appendix, a similar approach appears 
possible for streamers. As discussed also in \cite{short}, building 
on such a reduced description of streamer dynamics
appears very desirable, not only because it might make numerical
studies much easier, but also because it will allow us to draw upon
the knowledge and methods which have been developed in the last decade
in the field of interfacial pattern formation and dynamics
\cite{reviews}. The first step towards this goal is to determine the
field dependence of the velocity and the ionization and charge profile
of a planar front which propagates into the non-ionized region. We
thus analyze in this paper the inner problem for a planar streamer
front.  This allows us to reduce the problem to effectively one
dimension.  Our analysis clearly identifies the problem of streamer
front propagation as an example of front propagation into unstable
states.  Physically, the instability of the non-ionized gas against
charge fluctuations can be traced back to the fact that any small
electron density gets amplified by the impact ionization. As is
standard for front propagation into unstable states
\cite{benjacob,vs1,vs2,vs3,Oono}, we find that the one-dimensional
streamer equations exhibit a one-parameter family of uniformly
translating front solutions, parametrized by their velocity. As usual
\cite{benjacob,vs1,vs2,vs3,Oono}, the question is then to decide which
of these front solutions is the dynamically selected one, i.e., is the
one reached at long times after a localized ionized region has been
created by some initial ionization event. The existing knowledge of 
front propagation into unstable states \cite{vs1,vs2} provides us 
with an educated guess for the selected velocity, which we confirm 
with the help of numerical studies. Taken together, our results provide 
an essentially complete solution of the inner problem of planar streamer 
fronts.

In itself, the idea to analyze the planar fronts of a streamer 
model is not new --- we refer to \cite{TO,AT,Fowler,Dya} for 
earlier work. Apart from the fact, that the authors from the seventies
\cite{TO,AT,Fowler} investigate different 
models, which are more inspired by equilibrium concepts (e.g.,
the ionization behind the front is determined by thermal ionization,
where the electron temperature is raised by application of strong 
electric fields), our work casts new light on this old problem
from two different angles:\\
First of all, it was empirically noted, that the standard
approach to analyze uniformly translating fronts, failed to determine
a unique propagation velocity, given the field and the gas parameters.
Turcotte and Ong \cite{TO} clearly state this failure of their
theory (this ``great defect'' of their theory is recalled in Fowler's
reviews \cite{Fowler}) and suggest, that a unique solution might be
determined by a dynamical stability analysis.
Albright and Tidman \cite{AT} then perform such a 
stability analysis, but not in a systematic way, and they draw
incorrect conclusions. D'yakonov and Kachorovskii \cite{Dya} also 
find the indeterminacy of the speed of uniformly translating planar
fronts, now for an approximated version of our model, and propose to 
solve this by using the tip radius of the streamer finger as an 
extra length scale, which they, however, cannot determine.
We, in contrast, trace the indeterminacy of the velocity
from the analysis of uniformly translating streamer front 
solutions to the fact, that this is an 
example of front propagation into unstable states. Applying the 
concepts explained above, we solve the selection problem for planar 
fronts without additional assumptions or approximations.
We argue that a particular front solution out of a whole family of
dynamically stable solutions is selected, because it is the 
only one compatible with the initial condition of a localized 
ionization seed.\\
Secondly, this result is the first {\em ingredient} for studying
the formation of patterns, in particular of the tip radius ---
we do not attempt to model {\em global features of the pattern 
formation} with our planar front analysis. 
Our approach thus is very different in spirit to
the earlier investigations: As also stressed in \cite{short},
in an effective interface description based on a matched asymptotic
expansion, the results of weakly curved, almost planar fronts
are essentially used {\em locally} everywhere in the interface region:
They enter the analysis on the outer scale as boundary conditions at
the moving interface. It is on this outer scale that pattern formation
problems like the size, velocity and shape of the streamer 
should be analyzed. Once our results on planar fronts will be extended 
to weakly curved fronts, all the necessary ingredients to tackle these
questions appear to be available.

The main results of our present analysis of the streamer equations used
in the simulations \cite{DW,Vit} can be summarized as follows:

{\em (a)} Dimensional analysis shows that in dimensionless units, a
single parameter remains to characterize the gas, the dimensionless
electron diffusion coefficient $D$ characteristic of the gas [see
Eq.~(\ref{2-8})]. For gases under normal conditions, $D$ is small,
of order 0.1-0.3.

{\em (b)} The length scale set by the electron impact ionization
coefficient [the coefficient $\alpha_0^{-1}$ in Eq.~(\ref{2006})] is
on the order of microns for nitrogen. For $D \lesssim 1$ the thickness
$\ell_{in}$ of the charged layer is on the order of this same
ionization length for {\em negatively charged streamer fronts} {\em
  (NSF)} \cite{nomenclature}. Given that typical streamer diameters
found in the simulations are of the order of $1$mm, $\varepsilon =
\ell_{in} / \ell_{out}$ is at most of order $ 10^{-2}$; this justifies
an effective interface description of streamer dynamics.

{\em (c)} We find that electron diffusion acts as a singular
perturbation for {\em positively charged streamer fronts (PSF)}:
without diffusion, such fronts can not propagate, but with any nonzero
$D$, they do. As a result, the behavior is singular in the limit $D
\rightarrow 0$: for $D={\cal O} (1)$, the thickness $\ell_{in}$ is
again of order of the ionization length, but for $D\rightarrow 0$ the
electron density and its gradients diverge due to the appearance of
another smaller length scale (of order $D/ \alpha_0 $).

{\em (d)} The electron density generated by the propagating front is
again basically set by dimensional analysis for {\em NSF}.  We
calculate for $D \lesssim 1.5$ the dependence of the dimensionless
electron density $\sigma^-$ behind the front on the electric field
$E^+$ far ahead of our planar front. Our results compare favorably
with those extracted from the simulations \cite{Vit}, according to the
prescriptions of the theory of matched asymptotic expansions
\cite{vDyke,bender}.  Namely, $E^+$ is {\em not} the field value at
the electrode position, but the value obtained by extrapolating the
slowly varying outer field to the front position. We also calculate
the full $D$ and $E^+$ dependence of the electron density $\sigma^-$
behind the front of {\em PSF} for $D \lesssim 1.5$.

{\em (e)} The dynamically relevant (``selected'') front velocity $v_f$
is a unique function of $E^+$ and $D$. The analysis confirms the
strong asymmetry between $NSF$ and $PSF$ also found in the simulations
\cite{DW,Vit} for fronts propagating into an essentially non-ionized
region. The asymmetry is the stronger the smaller $D$ and disappears 
for $D \gg 1$.

{\em (f)} For $NSF$, $v_f$ is given by the so-called linear marginal
stability velocity $v^*$ \cite{vs1} --- see Eq.~(\ref{w1}) below. 
For parameter values used in the simulations, we find that $v_f$ is 
typically 30 to 40 \% higher than the electron drift velocity just 
in front of the streamer head, which agrees semi-quantitatively 
with the findings of Vitello {\em et al.} \cite{Vit}.

{\em (g)} We find that $PSF$ propagate for any nonzero value of the
dimensionless electron diffusion coefficient $D$. Due to the singular
behavior as $D \rightarrow 0$, we find that fronts propagate with a
unique velocity $v^\dagger$ predicted by the so-called nonlinear
marginal stability mechanism \cite{vs2} for small $D$. For the Townsend
expression used in the simulations \cite{DW,Vit}, this happens below a
well-defined field-dependent value of $D$ of order unity [see Fig.~3].
Above this threshold value, $PSF$ propagate with the linear marginal
stability value $v^*$.

In this paper, our main focus will be on those results that are of
greatest interest from the point of view of understanding the
generation of low temperature plasmas by the streamer mechanism. We
note, however, that the equations for planar streamer fronts
[Eqs.~(\ref{308}) and (\ref{309}) below] appear to be of interest in
their own right. As will be discussed briefly in Section V, our
streamers have several features in common with the celebrated
nonlinear diffusion equation studied in mathematics \cite{aw,eckmann}
since the early work of Kolmogorov {\em et al.} \cite{kpp} and Fisher
\cite{fisher}; at the same time, however, they are sufficiently more
complicated that they appear to present new challenges from a
mathematical point of view.

This paper is organized as follows. In Section II we introduce the
basic equations for streamer formation, and perform a dimensional
analysis for the inner problem of streamer fronts. In Section III, we
discuss the stability of the basic homogeneous states of interest, the
homogeneous non-ionized state and the homogeneous weakly ionized
state. We also discuss the physical mechanism of streamer formation
and the proper initial and boundary conditions to study these in the
case of planar fronts, which allow us to simplify the equations
describing planar front dynamics. In Section IV we demonstrate that
there exists a one-parameter family of uniformly translating fronts 
characterized by a continuous range of front velocities $v$.
We also briefly show how in the case $D=0$, the equations for
uniformly translating fronts can be solved analytically. These
solutions, which turn out to be useful as a small-$D$ approximation
for $NSF$, yield an explicit expression for the electron density
$\sigma^-$ behind the $NSF$ in terms of the field $E^+$ just ahead of
it. This is followed by an analysis of the general case $D \neq 0$;
then the equations can not be solved analytically, but we demonstrate
that there still is a one-parameter family of uniformly translating
front solutions. For $PSF$, we show that the limit $D \rightarrow 0$
is singular; we discuss this limit in detail and show that it accounts
for the strong asymmetry between $PSF$ and $NSF$ for realistic values
of $D$. In Section V we then summarize some of the main results 
\cite{benjacob,vs1,vs2,vs3,Oono} concerning the so-called selection 
problem, the question which particular front solution from the family 
is reached asymptotically for large times for a large class of initial
conditions. Application of these concepts allows us to predict the
shape and velocity of the dynamically relevant front solution (the
selected front) and the value of the electron density generated behind
it. This yields the various selection results for $NSF$ and for $PSF$,
summarized in points {\em (c)-(g)} above, and leads us to predict that
the behavior of $PSF$ in the limit $D\rightarrow 0$ is singular.  In
Section VI we present numerical simulations of the full partial
differential equations for planar streamer dynamics; starting from
various initial conditions, we illustrate, that in all cases we have
studied the long time dynamics of the system is characterized by a
$NSF$ and a $PSF$ whose behavior is in full agreement with our
predictions.  In the concluding section we finally reflect on our
results and on the future steps to be taken to arrive at an effective
interface description of streamer dynamics. In an appendix we discuss
differences and similarities between combustion and streamer fronts.

\section{Modeling and Dimensional Analysis}

\subsection{The Minimal Streamer Model }

For simulating the dynamical development of streamers out of a
macroscopic initial ionization seed in a so-called non-attaching 
gas like N$_2$ under normal conditions, Dhali and
Williams \cite{DW} and Vitello {\em et al.} \cite{Vit} use the
following set of deterministic continuum equations for the
electron density $n_e$, the ion density $n_+$ and the electric 
field ${\cal E}$: balance equations for electrons and ions,
\begin{eqnarray}
\label{2011}
\partial_t\;n_e \;+\; \nabla_{\bf R}\cdot{\bf j}_e
&=& source ~,
\\
\label{2012}
\partial_t\;n_+ \;+\; \nabla_{\bf R}\cdot{\bf j}_+
&=& source ~, \end{eqnarray}
where the fact that the two source terms are the same is due to charge
conservation in an ionization event; the Poisson equation, 
\begin{equation}
\label{2013}
\nabla_{\bf R}\cdot {\cal E} = {{e}\over{\varepsilon_0}} \;(n_+ -n_e)
~, \end{equation} and the approximate phenomenological expressions
\begin{eqnarray}
\label{2014}
{\bf j}_e &=& - n_e \;\mu_e \;{\cal E} - D_e \;\nabla_{\bf R} \;n_e ~,
\\
\label{2005}
{\bf j}_+ &=& 0~,
\\
\label{2006}
source &=& |n_e \mu_e {\cal E}| \;\alpha_0
\;\mbox{\large{e}}^{\textstyle -E_0/|{\cal E}|}~.  \end{eqnarray} 
Apart from the
fact that we will allow for a slight generalization of
Eq.~(\ref{2006}), these are the equations that we will investigate
analytically below.

In these equations, ${\bf j}_e$ and ${\bf j}_+$ are the particle
current densities of electrons and positive ions, and $e$ is the
absolute value of the electron charge. The (dimensional) spatial
coordinates are denoted by $\bf R$, and $\nabla_{\bf R}$ is the
gradient with respect to these coordinates.  The use of only
Poisson's law of electrostatics, Eq.\ (\ref{2013}), means that all
magnetic fields as well as terms in the Maxwell equations associated
with time-dependences of the fields, are neglected \cite{maxwell}.

The electron particle current density ${\bf j}_e$ is approximated in
(\ref{2014}) as the sum of a drift and a diffusion term.
Note that this diffusion approximation implies that the electron 
mean free path must be small with respect to the scale of 
variation $\ell_{in}$ of the electric field. This condition is 
just about satisfied for the parameter values taken for
$N_2$ in the simulations, except possibly at the highest field
values (see also the discussion in the concluding Section VII).
The electron drift velocity is taken to be linear in the field 
${\cal E}$, with $\mu_e$ the (positive) electron mobility. 
The electron diffusion coefficient $D_e$ and the mobility 
$\mu_e$ are treated here as independent coefficients,
since they effectively depend on the field strength \cite{Raizer}
(only in the low-field limit are they related by the Einstein
relation). More generally, the diffusion coefficient should be 
replaced by a diffusion tensor, which is diagonal in a
reference frame with one axis along the electric field. Its
longitudinal component, the only relevant one for planar fronts
perpendicular to $E$, is somewhat smaller than the transverse one.
Since we will see, that N$_2$ reaches a typical degree of
ionization of only $10^{-5}$, density fluctuations of the non-ionized
gas can be neglected and the mean free path of the electrons and
therefore $\mu_e$ and $D_e$ can be taken as independent of the degree
of ionization.

The ionic current is neglected according to Eq.\ (\ref{2005}), 
since the mobility of ions is at least two orders of magnitude 
smaller than that of the electrons \cite{DW}. In particular, for 
the analysis of the inner scale, that we will perform in the 
present paper, ${\bf j}_+$ is negligible.

The $source$ (\ref{2006}) finally accounts for the creation of free
charges by impact ionization. If the product of electric field ${\cal
  E}$ and electronic mean free path $\ell_{mfp}$ is large enough, free
electrons can gain sufficient kinetic energy to ionize neutral
molecules.  Accordingly there is a threshold field $|{\cal E}|=E_0
\propto {\ell_{mfp}}^{-1}$. For $|{\cal E}| \gtrsim E_0$ the
probability that a scattering event carries at least the ionization
energy is large.
The effective ionization cross-section $\sigma_{cs}(|{\cal E}|)$ then
essentially saturates, while for $|{\cal E}| \ll E_0$ the ionization
rate per scattering event is largely suppressed. The $source$-term is
given by the ionization rate, which can be calculated as the product
of the drift current of free electrons $|n_e\mu_e {\cal E}|$ times the
target particle density $n_n$ of the neutral gas times the effective
ionization cross-section $\sigma_{cs}(|{\cal E}|)$.  Commonly, a
phenomenological ionization coefficient $\alpha(|{\cal E}|) = n_n
\sigma_{cs}(|{\cal E}|)$ is used, (which clearly has dimension of
inverse length,) whose field threshold behavior in the Townsend
approximation $\alpha(|{\cal E}|)=\alpha_0 \exp(-E_0/|{\cal E}|)$
\cite{Raizer} is expressed by Eq.\ (\ref{2006}). As discussed by
Raizer \cite{Raizer}, in the approximation that every collision is
ionizing, if the electron carries an energy larger than the 
ionization energy $I$, we have
\begin{equation}
\label{mfp}
\alpha_0 \approx \ell^{-1}_{mfp}~,\quad \mbox{and } \quad E_0 \approx
I/(e\;\ell_{mfp})~.
\end{equation}
Since in much of our analysis the specific form of $\alpha(|{\cal
  E}|)$ is not needed, we will use a slightly more general formulation
in Eq.\ (\ref{2022}) below. 

In the $source$-term, ionization due to the photons also created in
recombination or scattering events, is neglected. This is motivated by
the ionization cross-sections due to photons being much smaller than
those due to electrons. Note that, if photo-ionization is taken into 
account, the dynamical equations become nonlocal.

No sink term needs to be included for the analysis of the inner
problem, since the recombination length at a degree of ionization of
order $10^{-5}$ that we will derive below is very large as compared
with the front width $\ell_{in}$. (For this reason, the inner 
problem is the same for streamers and leaders \cite{Raizer}: 
the difference between these discharge modes, which consists 
in the fact that recombination is non-negligible in the plasma 
body of leaders, would come into play only when solving, 
at a later stage, the outer problem.) The fact that the
degree of ionization remains small is also the reason that 
saturation effects are neglected in (\ref{2006}).

In contrast to the situation in N$_2$, that is described by
our model equations, in attaching gases like O$_2$, a third charged
species plays a role, namely negative ions formed by a neutral
molecule catching a free electron. For a description of the physics
of such attaching gases and simulations thereof, see, e.g., 
\cite{nonattach}.

The equations above are deterministic.
Thermal fluctuations in fact can be neglected, since
even an unphysically small ionization energy of 3 eV leads 
to a Boltzmann factor of $10^{-52}$ at room temperature.
Also other stochastic effects are not accounted for in the 
simulations we compare to. We further discuss possible 
stochastic effects in the experiments in the conclusion.

Finally, the dynamical system (\ref{2011})-(\ref{2006}) must be
complemented by:

{\em (i)} boundary conditions: as will be discussed in detail in
Section III, for the problem of front propagation, these are specified
by the value $E^+$ of the electric field far ahead of the front, where
the total charge density vanishes.

{\em (ii)} initial conditions: we ignore the details of the plasma
nucleation event (e.g. triggering by radiation from an external
source), and assume that at $t=0$ a small well-localized ionization
seed is present. The precise meaning, for our problem, of
``well-localized'' will be made clear in Section V.

\subsection{Dimensional Analysis}

In order to identify the physical scales and intrinsic parameters of
our problem, we reduce Eqs.~(\ref{2011})-(\ref{2006}) to a dimensionless
form. The most natural scale of length and electric field are the
ionization length $\ell_{ion}= \alpha_0^{-1}$ and the threshold field
$E_0$ of the ionization rate (\ref{2006}). The velocity scale is then
the electron drift velocity at this field strength, $v_0=\mu_e E_0$,
leading to a time unit $t_0=(\alpha_0 \mu_e E_0)^{-1}$, and a charge
unit $q_0=\varepsilon_0 \alpha_0 E_0$.

For concreteness, we list here the values of these quantities for
N$_2$ at normal pressure, used in the simulations \cite{DW,Vit} 
\begin{eqnarray}
&\alpha_0^{-1} \approx 2.3\;\mu\mbox{m}~ , \quad 
& \quad v_0 \approx 7.56 \cdot 10^7\;\mbox{cm/s}~, \nonumber\\ 
&t_0 \approx 3\cdot10^{-12} \;\mbox{s}~, \quad 
& \quad q_0 \approx 4.7 \cdot10^{14}\; e/\mbox{cm}^3~    \label{2026}   \\
\nonumber &E_0 \approx 200 \;\mbox{kV/cm} \quad 
& \quad \mu_e \approx 380 \;\mbox{cm}^2\mbox{/Vs}~.  
\end{eqnarray} 
We now introduce dimensionless quantities by defining 
\begin{equation}
\begin{array}{ll}
  \nonumber {\bf r}={\bf R}\; \alpha_0 ~, & \tau= t \;/t_0 ~, \\ 
   q=(n_+ - n_e)\; e/ q_0
  ~,\hspace*{8mm} & \nonumber \sigma= n_e \; e /q_0 ~, \\ 
\label{2-7} {\bf j} = -{\bf j}_e \;
e / (q_0v_0) ~, & {\bf E} = {\cal E} /E_0 ~.
\end{array}
\end{equation} 
Note that with our definition, ${\bf j}$ now plays the role of a
dimensionless {\em charge} current. If we furthermore introduce the
dimensionless diffusion coefficient $D$ as \begin{equation}
\label{2-8}
D= D_e \alpha_0 / \mu_e E_0~, \end{equation} 
we obtain what we call the streamer equations 
\begin{eqnarray}
\label{2022}
\partial_\tau\;\sigma \;-\; \nabla\cdot{\bf j}
&=& \sigma \; f(|{\bf E}|)~,
\\
\label{2023}
\partial_\tau\;q \;+\; \nabla\cdot{\bf j}
&=& 0~,
\\
\label{2024}
q &=& \nabla\cdot{\bf E}~,
\\
\label{2025}
{\bf j} &=& \sigma \;{\bf E} + D \;\nabla \sigma~, \end{eqnarray} 
where $\nabla$ denotes the gradient with respect to the 
dimensionless coordinate {\bf r}, and  where the ``ionization 
function'' \begin{equation}
\label{2-13}
f(|{\bf E}|) = |{\bf E}| \; \alpha(|{\bf E}|) / \alpha_0~ 
\end{equation}
is assumed to vanish at zero field. Townsend's expression 
(\ref{2006}) yields: 
\begin{equation}
\label{2-14}
f_T(|{\bf E}|) = |{\bf E}| \; \exp(-1/ |{\bf E}|)~.  \end{equation} 
In general,
we will treat an ionization function with the properties
\cite{footnotef} \begin{equation}
\label{2-17a}
f(0)=0=f'(0)~,\quad \mbox{and} \quad f'(|{\bf E}|) \ge 0
\quad \mbox{for all }~|{\bf E}|~.  \end{equation} The dimensionless
equations (\ref{2022})-(\ref{2025}) now depend on only one internal
parameter, the dimensionless diffusion coefficient $D$.  For the
values used in \cite{DW,Vit} for N$_2$
under normal conditions, $D\approx 0.1$, while according to the data
given by Raizer \cite{Raizer}, for Ne and Ar, $D\approx 0.3$. We
believe that typical values are generally in the range 0.1-0.3, since
in the approximation (\ref{mfp}), $\alpha_0/E_0 \approx I/e$ and since
the ratio $D_e/ \mu_e$ appears to be commonly of the order of volts
for large fields, while $I$ is typically of the order of several
electron volts.

We are now able, solely on the basis of the dimensional analysis
above, to make a first semi-quantitative prediction about streamers.
We will in practice be interested in external fields $E^+ = {\cal
  O}(1)$ (for $E^+ \ll 1$ and $\alpha_0^{-1}$ on the order of
micrometers, the electron avalanche process becomes much too
ineffective for streamer fronts to develop at reasonably small
distances; also our scale separation approach discussed in the 
introduction might break down). We can therefore expect that, for $D$ 
values $\lesssim 1$, as is the case for N$_2$, front widths will be 
of order $\alpha_0^{-1}$, and that in addition the reduced electron 
density $\sigma^-$ far behind the front on the inner scale will be of 
order unity as well. This leads one to expect electron densities in 
the streamer body on the order of $10^{14}$ cm$^{-3}$, in agreement 
with numerical findings.

\section{Homogeneous Solutions and the Concept of Fronts}

\subsection{Homogeneous States and their Stability}

The first task, when studying in general the propagation
of a front, is to identify the nature and stability of the 
states which the front connects. We expect the invading state, 
here the ionized one created by the front, to be stable 
\cite{footnote5a}, while the invaded state can in general 
either be metastable or unstable. Physically, we of course 
expect the non-ionized state to be unstable in a non-vanishing
field in the present model. (In an attaching gas 
forming also negative ions, it is conceivable that the 
non-ionized state is metastable for not too strong fields.)

Equations (\ref{2022})-(\ref{2025}) immediately yield, 
that stationary homogeneous states simply are solutions of 
\begin{equation}
\label{2-15}
\sigma f(|{\bf E}|) =0~.  
\end{equation} 
So, these stationary states decompose into two families:

{\em (i)} Non-ionized states, with $\sigma=0$, ${\bf E}$ arbitrary:
Since the density of free electrons vanishes, no ionization can occur,
whatever the value of ${\bf E}$ is. If also the density of ions
vanishes, $\nabla \cdot {\bf E}=0$.  Since these states correspond to
the physical situation far ahead of the front, we label them (+).
Moreover, since we will need in particular the case in which the field
ahead of the front is constant, we take ${\bf E}^+=$const.

{\em (ii)} Completely screened states, labeled ($-$), with ${\rm
  E}=0$, $\sigma^-$ arbitrary \cite{footnote5ab}: Whatever the
electron density, for ${\bf E}=0$ impact ionization does not occur and
thermal energy is much too small to permit ionization.

Since the steady states we consider as well as the equations of motion
are translation invariant in space and time, the eigenstates of the
linear perturbations are Fourier modes of the form 
\begin{equation}
\left( \begin{array}{l} \delta \sigma ({\bf r},t) \\ 
                        \delta {\bf E}({\bf r},t)
       \end{array} \right) 
= \left( \begin{array}{l} \sigma_1 \\ {\bf E}_1 \end{array} \right) 
\exp(i{\bf k} \cdot {\bf r} + \omega \tau)~.
\label{2-16}
\end{equation} 
We first investigate the linear stability of the non-ionized state
$\sigma^+=0$. Upon linearizing the equations about the zeroth order
values $(\sigma^+=0, {\bf E}^+)$, we find two branches of modes:

{\em (a)} The first, trivial branch is a zero mode $(\omega=0)$, with
$\sigma_1=0$, expressing that the electron density remains zero. This
zero mode accounts for the degeneracy of the non-ionized states, i.e.,
for the fact that there exists a (+) stationary state for each value
of ${\bf E}^+$. (For ${\bf E}^+ \neq {\rm const.}$, these zero modes
express the degeneracy of all steady states with $q^+=\nabla \cdot
{\bf E}^+$ for any ion density $q^+$ as long as the electron density
$\sigma^+$ vanishes.)

{\em (b)} The second branch of perturbations is associated with
fluctuations carrying a finite electron charge; its dispersion
relation is \begin{equation}
\label{2-17}
\omega^+ = i {\bf k}\cdot {\bf E}^+ + f(|{ \bf E}^+|) -Dk^2~, 
\end{equation} with
$i\omega^+ {\bf k} \cdot {\bf E}_1 =(f(|{\bf
  E}^+|)-\omega^+)\sigma_1$. The first term on the r.h.s. of
(\ref{2-17}) simply expresses the fact that the electrons drift, to
first order, in the electric field ${\bf E}^+$ with velocity $(- {\bf
  E}^+)$. The real part $\Re \omega^+$, the sign of which determines
whether fluctuations decay or are amplified, contains a destabilizing
term, expressing that any small electron density fluctuation is
amplified at rate $f$, and a stabilizing term, due to the diffusive
spreading of electron charges.  For $k^2<f(|{\bf E}^+|)/D$, $\Re
\omega^+>0$: non-ionized states are unstable against long-wavelength
perturbations.

We note, that the single Fourier eigenmodes (\ref{2-16}) violate
individually the physical constraint that $\sigma$ be positive
everywhere. But Eq.~(\ref{2-17}) also determines the time evolution 
of physically allowed fluctuations (wavepackets) that are 
superpositions of these eigenmodes. For example, one easily deduces 
from it Lozanski's expression \cite{Loz} for the time-evolution of 
a Gaussian-shaped small electron density with arbitrary constants
$c_1,c_2>0$, 
\begin{eqnarray}
\label{2034}
\delta \sigma({\bf r},\tau) &=& c_1\;
\mbox{\large e}^{\textstyle \;f(|{\bf E}^+|)\:\tau }\;\;
(c_2+4D\tau)^{-3/2} \nonumber\\
&& \qquad\mbox{\large e}^{\textstyle -\;
({\bf r}+{\bf E}^+\tau)^2/(c_2+4D\tau) }~, 
\end{eqnarray}
as long as linearization around the non-ionized state holds. 
As expected, the center of the spreading packet drifts with 
velocity $-{\bf E}^+$, while the total number of electrons it 
contains is amplified at rate $f$ and the wave-packet
stays Gaussian, with time-dependent width $c_2+4D\tau$. 
Such ionization modes derived by linearizing around the 
non-ionized state are known as electron or ionization avalanches 
in the gas discharge literature.

We now perform the same linear stability analysis for the completely
screened states $(\sigma^-={\rm const.},{\bf E}^-=0)$. The fact that 
$f'(0)=0$ from Eq.\ (\ref{2-17a}) assures that the linear 
perturbations are not affected by ionization; the dynamics thus 
evolves with conserved particle densities.

Again, due to the existence of a continuous family of screened
stationary states, parametrized by $\sigma^-$, the spectrum contains a
branch of $\omega=0$ modes.  For the nontrivial branch, the dispersion
relation is given by \begin{equation}
\omega^- = -\sigma^- -Dk^2~,
\label{2-19}
\end{equation} 
while the eigendirection of such a perturbation is given by 
\begin{equation}
\label{2-20}
\sigma_1 + i {\bf k}\cdot {\bf E}_1 =\sigma_1 + q_1 =0~.  
\end{equation}
Since
$(\sigma_1+q_1)$ is the dimensionless ion density of the linear mode,
Eq.~(\ref{2-20}) simply expresses the fact that ions are completely
immobile in our model.

Equation (\ref{2-19}) expresses the fact that the completely screened
($-$) states are stable, the decay of perturbations being due to the
added stabilizing effects of overdamped plasmons $(-\sigma^-)$ and
electron diffusion. The ${\bf k}\to 0$ limit of the plasmon
mode leads to dielectric screening \cite{panof}.

\subsection{The Mechanism of Front Creation}

Let us now investigate the dynamical evolution of an initial state in
which the electron and ion densities vanish everywhere except in a
small localized region. An example of such localized initial
conditions is an initially Gaussian electron density, as in the
simulations \cite{DW,Vit} --- under what circumstances initial
conditions are sufficiently localized will become clear later.  As
long as the electron and ion densities are small enough, we can
neglect in linear approximation the changes in the field as we did
above when linearizing about the non-ionized state.  As a result, both
densities will grow due to impact ionization. If this were the only
mechanism, the space charge would remain unchanged and the ionization
would continue indefinitely. However, the electrons are mobile, and at
the same time they start to drift in the direction opposite to the
electric field $ {\bf E}$. If we neglect for the moment the diffusion,
this drift has two effects: First of all, the electrons start to drift
in the direction of the anode. Impact ionization then starts in
previously non-ionized regions as well, so the ionized region expands
towards the anode. Secondly, as the electrons drift while the ions
stay put (on the fast time scale), a charge separation occurs which
tends to suppress the field strength in the ionized region. When the
size of the initial perturbation and/or the time during which the
avalanche has built up are large enough, the screening of the field
becomes almost complete in the ionized region so that ionization stops
there. The behavior in this region can be described
by linearizing around the screened state as done above. After an
electrically screened body of the ionized region has developed, the
initial ionization avalanche is said to have developed into a
streamer. Thus streamer fronts are strongly nonlinear and determined
by two competing mechanisms, which dynamically balance each other: the
ionization process which is strongest at the leading edge and the
screening of the field due to the free charges which increases towards
the rear end of the front.  This balance also explains our finding
that the ionization length and the screening length in the plasma
behind the $NSF$ are of the same order of magnitude for field values
that are not too small. Technically speaking, the challenge in
constructing the full front is to connect the two regimes linearized
about the homogeneous states in an appropriate way through the
nonlinear regime of the front.

In the above discussion, we have neglected electron diffusion. In this
case the $NSF$ propagates towards the anode with a velocity that is at
least the drift velocity of the electrons in the local electric field.
The $PSF$, in contrast, is moving in the direction {\em opposite} to
the drift of the only mobile species, the electrons. Its space charge
is formed by the ions staying put, while the electrons are drawn into
the ionized body. Propagation of a
$PSF$ is therefore only possible if the electron diffusion current
overcompensates the drift current. This in turn implies that if the
diffusion coefficient $D$ is small, electron density gradients must be
extremely steep. From this discussion it already becomes evident ---
and we will derive this below --- that for an $NSF$, diffusion is a
small correction for $D \ll 1$, since drift and diffusion currents are
acting in parallel directions. In $PSF$, however, diffusion has to
overcome the drift, and as a result in this case the limit of
vanishing diffusion is very singular. We will see in Section IV that
this manifests itself through the emergence of a new inner length
scale $D/ \alpha_0 = D_e/(\mu_e E_0)$, the diffusion length associated
with the electron drift velocity.

Of course, a charged front only screens the normal component 
of the electric field. This is why electric screening is efficient 
in the head of the streamer, while the field penetrates in
the body of a single streamer in the simulations \cite{DW,Vit}.
Our planar front analysis thus serves as a first approximation for
the mechanisms in the moving tip of the streamer finger.

\subsection{The one-dimensional Streamer Equations}

Let us now restrict our analysis to the case of plane fronts
perpendicular to a constant electric field. Of course, in practice
planar streamer fronts will be unstable to deformations along the
front (very much like in the Mullins-Sekerka instability in crystal 
growth \cite{reviews}), but as explained in the introduction, the 
planar front analysis is a first step towards understanding the 
dynamics on the inner length scale $\alpha_0^{-1}$ and time scale 
$t_0$. As such, it is the first basic ingredient for deriving an 
effective interface model on scales $\gg \alpha_0^{-1}$.

We choose the $x$ axis as parallel to the field and perpendicular to
the planar front so that $ {\bf E} = E \hat{\bf x}$ and $ \nabla =
\hat{\bf x}\;\partial_x$.  From the point of view of matched
asymptotic expansions, the electric field in the non-ionized region
before the front will vary adiabatically slowly on the ``inner'' time
scale $\tau$ of the front, the timescale on which the front propagates
over a distance comparable to its width, because the length scales of
the outer problem determining the changes of $E$ are assumed to be
much larger than the inner scale $\alpha_0^{-1}$.  For our study of
the inner problem, we thus take the asymptotic field value ${\bf E}^+$
in the unionized region constant in time.  Furthermore, we will use
the convention that the unionized initial state into which the front
propagates is at the right towards large positive values of $x$, so
that there 
\begin{eqnarray}
\label{304}
\label{2041}
\sigma \to \sigma^+ = 0~,\;\; q \to q^+ = 0~,\;\; &&E \to E^+~~, 
\quad \partial_{\tau} E^+=0~, 
\nonumber\\
&&\mbox{for } x \to +\infty~, 
\end{eqnarray}
which motivates now the use of the superscript +.  We emphasize again
that ``$x \to +\infty$'' should be interpreted on the length scale
$\alpha_0^{-1}$ of the inner problem in the sense of matched
asymptotic expansions \cite{vDyke,bender}. Far behind the front, i.e.,
for $x \to -\infty$, the discussion of Section II leads us to expect a
homogeneous stable state \begin{eqnarray}
\label{305}
\label{2042}
\sigma \to \sigma^- \ne 0~,\quad q \to q^- = 0~,\quad &&E \to 0~, 
\nonumber\\
&&\mbox{for } x \to - \infty~.  
\end{eqnarray} 
Which value $\sigma^-$
will be dynamically selected and what the corresponding front velocity
and profile are, for a given fixed value of the electric field $E^+$
before the front, is the selection problem, we aim to solve.

The boundary condition (\ref{304}) allows an important simplification
of the equations in one dimension: If we insert (\ref{2024}) into
(\ref{2023}), we obtain \begin{equation}
\label{306}
\partial_\tau\;E + j \;=\; h(\tau)~,
\end{equation}
where $h(\tau)$ is an arbitrary function of time which is constant
in space.  In view of the boundary condition
(\ref{304}), $h(\tau)$ vanishes at $x \to \infty$ and thus everywhere.
For planar fronts, the model Eqs.\ (\ref{2022})-(\ref{2025}) then
reduce to \begin{eqnarray}
\label{307}
\partial_\tau\;\sigma 
&=& \partial_x\;(\sigma \;E) \;+\; D \;{\partial_x}^2 \sigma \;+\;
\sigma \; f(|E|)~,
\\
\label{308}
\partial_\tau\;E
&=& \;-\;\sigma\;E\;-\;D\;\partial_x\; \sigma~, \end{eqnarray} 
with space charge and electric current given by \begin{equation}
\label{309}
q\;=\;\partial_x\; E \qquad\mbox{and}\qquad
j\;=\;\sigma\;E\;+\;D\;\partial_x \;\sigma~.  
\end{equation}
We will refer to this set of equations as to {\em the one-dimensional 
streamer equations}. They are the basic equations of this paper, on 
which the rest of our analysis will be based.

Eq.~(\ref{308}) implies that the field decays behind the front, if no
strong density gradients act against it. As we shall see later when we
will discuss our simulation results in Section VI, such strong density
gradients often occur during the transient regime before a $PSF$
emerges. Once, however, a front has approached an approximately
uniformly translating state, the electron density $\sigma^-$ behind
the front is almost homogeneous and the field behind the front then
decays to zero on a time scale $1/ \sigma^-$ according to (\ref{308}).
Note, that the local decay of the field for any nonzero electron 
density is due to electrodynamics of conserved quantities that 
continues also after the impact ionization has been suppressed.

We finally note that in the limit where the diffusion is small ($D \ll
1$), it is easy to identify the crossover time from the linear avalanche
regime to that of streamer propagation in the case that the initial
electron density is small and nonzero only in a very narrow localized
region. As explained in the beginning of this section, in the
avalanche regime we can neglect the changes in the background field
$E^+$ due to the build-up of the charges. The evolution of the electron
density is then described by the linearized version of (\ref{307}), a
linear equation with drift, diffusion and growth. Hence, if the
initial electron density is, e.g., Gaussian, the electron density will
according to (\ref{2034}) remain a Gaussian profile, whose maximum
drifts with a velocity $|{\bf E}| $ in the direction opposite to the field
and whose amplitude grows exponentially as $\exp(f(E^+)\tau)$ . In
other words, if the total initial electron charge is $N_e(0) = \int dx
\; \sigma(x,0)$, then the total number of electrons in this avalanche
regime grows as $N_e(\tau)=N_e(0) \exp(f(E^+)\tau)$. Likewise, the
total ion charge grows exponentially, but if both the diffusion constant
and the width and amplitude of the initial perturbation are small, 
the electron drift will separate the negative electron charge and the 
positive ion charge almost completely. The crossover to the nonlinear 
streamer regime will therefore occur when the total charge in the 
positively and negatively charged regions is big enough that screening 
of the field becomes appreciable, i.e., at a time $\tau_c$ when 
\begin{equation}
\label{309a}
N_e(\tau_c) \approx |E^+|~,~~~\Rightarrow~~~ \tau_c \approx
{{1}\over{f(E^+)}} \ln[|E^+|/N_e(0)]~.  \end{equation}

\section{Uniformly Translating Front Solutions}

Above we already have introduced the idea, that fronts asymptotically
approach some shape, which is independent of the initial conditions.
This is based on our experience \cite{benjacob,vs1,vs2,vs3,Oono} with
other examples of front propagation into unstable states that the
front will acquire some asymptotic shape and velocity in the long time
limit, which will be the same (``universal'') for a large class of
``sufficiently localized'' initial conditions that comprise most
physically relevant initial states. This property is often referred to
as the front selection problem.  Our subsequent analysis will
therefore follow the usual strategy in examples of this type: We will
first show in this section that there generally is a one-parameter
family of front solutions. In Section V we then summarize our present
understanding of the front selection problem, and on the basis of this
predict the properties of the selected streamer front.  The numerical
simulations that confirm our predictions are presented in Section VI.

Uniformly translating fronts with velocity $v$ are stationary in a
coordinate system moving with velocity $v$. If we denote this comoving
coordinate by $\xi=x-v\tau$, the partial differential equations ({\em
  pde's}) (\ref{308}) and (\ref{309}) in this coordinate system become
\begin{equation}
\begin{array}{rcl}
\label{3011}
\left.\partial_\tau\;\sigma\right|_\xi &=& v \;\partial_\xi\;\sigma
  \;+\; \partial_\xi\; (\sigma\;E) \;+\; D\;{\partial_\xi}^2\sigma
  \;+\; \sigma\;f(|E|)~ ,
\\
\left.\partial_\tau\;E\right|_\xi &=& v\;\partial_\xi\;E\;-\;\sigma\;E
  \;-\; D\;\partial_\xi\;\sigma ~.
\end{array}
\end{equation}
A front translating uniformly with velocity $v$ in the {\em fixed}
frame $x$ is stationary in this comoving frame, 
$\left.\partial_\tau\;\sigma\right|_\xi \;=\; 0 \;=\;
\left.\partial_\tau\;E\right|_\xi$.  
As a result, the corresponding front profiles are solutions 
of the ordinary differential equations ({\em ode's}).
(We continue to use partial differential signs
$\partial_\xi$ even though the uniformly translating solutions are
functions of the variable $\xi$ only.)
\begin{equation}
\begin{array}{rcrcrcccc}
\label{3013}
D\;{\partial_\xi}^2\sigma &+& (v+E)\;\partial_\xi \sigma &+&
\sigma\;\partial_\xi E &+& \sigma\;f(|E|) &=& 0~,
\nonumber \\
& & D\;\partial_\xi \sigma &-& v\;\partial_\xi E &+& \sigma \;E &=&
0~,
\nonumber \\
\end{array}
\end{equation}
These equations are analyzed below. Both for $D=0$ and for $D\neq
0$, they admit solutions for a range of values of the velocity, so we
are indeed faced with the question of front selection.

It is important to realize that not all the exact uniformly
translating front solutions of these {\em ode's} correspond to
physically relevant solutions. In particular, any physical electron
density $\sigma$ needs to be non-negative ($\sigma \geq 0$), but as we
shall see the set (\ref{3013}) admits $PSF$ solutions where $\sigma$
goes negative. We expect these solutions to be unstable (in accord
with the ``nonlinear marginal stability'' scenario \cite{vs2}), and
also not te be approachable from an initial condition with $\sigma \ge
0$. Hence they are neither dynamically nor physically relevant.
Furthermore, note that in our model the ion density $q_i$
($=\rho+\sigma$) can only increase due to impact ionization
[Eqs.~(\ref{2022}) and (\ref{2023}) imply $\partial_{\tau} q_i =
\sigma f(E) \geq 0$]. With our convention that the non-ionized state
is on the right, this implies that uniformly receding front solutions
with $v \leq 0$ are unphysical. We will therefore call a uniformly
translating front solution physical if
\begin{equation}
\label{3015}
v >0 \quad \mbox{and} \quad \sigma(\xi) \ge 0 \quad \mbox{for all}~\xi~.
\end{equation}

\subsection{$D=0$ Front Solutions}

In contrast to the case $D \neq 0$, where we can derive properties of
uniformly translating fronts only either qualitatively by discussing
flows in phase space or quantitatively by numerical integration, Eqs.\ 
(\ref{3013}) for $D=0$ can be integrated explicitly. Doing so, we
derive a simple explicit expression for the electron density
$\sigma^-$ behind the front in terms of the field $E^+$ before the
front; this analysis generalizes an earlier result of D'yakonov and
Kachorovskii \cite{Dya}, and explicitly illustrates the existence of a
family of uniformly translating solutions. For $NSF$, these results
extend smoothly to the case $D\neq 0$: The electron density
$\sigma^-(E^+)$ derived for $D=0$ will turn out to be a good
approximation for $D\lesssim 1$, and the small overshoot of $\sigma$
at the rear end of the front visible in the three-dimensional
simulations in Fig.\ 1(c), is also recovered for $D=0$. For $PSF$, on 
the other hand, we will see that $D$ acts as a singular perturbation, 
so that the class of $D=0$ $PSF$-solutions that we derive here is not
relevant for the $PSF$ selection problem for $D\lesssim 1$.

The {\em ode's} describing uniformly translating fronts for vanishing
diffusion are found by putting $D=0$ in (\ref{3013}).  These equations
then become \begin{eqnarray}
\label{401}
\partial_\xi\;\Big[ \;(v+E)\;\sigma\;\Big] &=& -\; \sigma\;f(|E|)~,
\\
\label{402}
v\;\partial_\xi\;\ln |E| &=& \sigma ~ .  \end{eqnarray} 
Upon insertion of the
l.h.s.\ of Eq.\ (\ref{402}) for $\sigma$ in the r.h.s.\ of Eq.\ 
(\ref{401}), this equation can then be expressed as a complete
derivative by writing \begin{equation}
\label{403}
\partial_\xi\;\left[ 
\;(v+E)\;\sigma \;+\; v\;\int_c^{|E|} dx\;\frac{f(|x|)}{x} \;\right]
\;=\; 0~.  \end{equation}
For physical fronts with $v > 0$ and $\sigma \ge 0$ [see
(\ref{3015})], we see from (\ref{402}), that $E$ is a monotonic
function of $\xi$, 
\begin{eqnarray}
\label{404}
\mbox{sgn}\;\Big(\partial_\xi E(\xi)\Big) \;=\; &&\mbox{sgn}\;q(\xi)
\;=\; \mbox{sgn}\;E(\xi) \;=\; \mbox{sgn}\;E^+ 
\nonumber\\
&&\mbox{for all }\xi~ .  
\end{eqnarray}
This allows us to use $E$ as a coordinate instead of
$\xi$. According to (\ref{2041}), before the front at $\xi \to \infty$
the electron density vanishes, so $\sigma^+= \sigma[E^+] = 0$.  Eqs.\ 
(\ref{403}) and (\ref{404}) together then determine $\sigma$ as a
function of $E$ as \begin{equation}
\label{406}
\sigma[E] \;=\; \frac{v}{v+E}\;\;\rho_{E^+}[E] ~, \end{equation}
with the
function \begin{equation}
\label{407}
\rho_{E^+}[E]\;=\; \int_{|E|}^{|E^+|} dx\;\frac{f(|x|)}{x} \;=\;
\rho_{|E^+|}[|E|] ~~(\geq 0)~.  \end{equation} 
The function $\rho_{E^+}[E]$ is nothing but the ion density,
as can be deduced by inserting $q = \partial_\xi E$ into 
(\ref{402}) and equating the charge density $q$ with
$\rho - \sigma$. The ion density $\rho$ for $D=0$ turns out
to be a function of $E$ and $E^+$ only, and to be independent 
of the particular front shape parametrized by $v$.

The fields $\sigma$, $\rho$ and $E$ as a function
of $\xi$ can be found by solving the implicit equation for $E=E(\xi)$
\begin{equation}
\label{408}
\partial_\xi\;\ln\;|E| \;=\; \frac{\rho_{|E^+|}[|E|]}{v+E}~,
\end{equation} 
which can be derived from Eqs.\ (\ref{402}) and (\ref{407}).

Eq.\ (\ref{406}) immediately shows that physically allowable solutions
with $\sigma \geq 0$ and $v>0$ must have $v+E \ge 0 $ for all field
values. Because of the monotonicity of $E$ as a function of $\xi$,
this is automatically satisfied for $PSF$ with $E^+>0$, but for $NSF$
this implies in particular that $v+E^+ \ge 0$; together with $v> 0$ we
thus have for physical fronts \begin{equation}
\label{409}
v \;\ge\; \mbox{max} \;\Big[\; 0\; , \;-E^+\; \Big]~.  \end{equation} 
In physical
terms, the condition $v \ge -E^+$ expresses, that the velocity of
uniformly translating fronts must be at least the drift velocity
$-E^+$ of free electrons in the leading edge of
the front, where the field is strongest. (Remember, that (\ref{404})
implies, that the field is monotonic in space.)

For all values of $v$ obeying the inequality (\ref{409}) the solutions
of (\ref{406}) and (\ref{408}) are proper, physically allowable
solutions for fixed $E^+$; within the context of the present model,
this illustrates a general feature of front propagation into unstable
states, namely that there exists a family of front solutions
parametrized by the velocity \cite{footnote6}.

In Fig.~2(a), we plot the solution (\ref{406}) for $\sigma$ as a 
function of $E$ for the fixed value of the velocity $v=2$ in the case 
that the impact ionization function $f(E)$ is given by the Townsend 
expression $f_T(E)$ of (\ref{2-14}) as in the numerical simulations
\cite{DW,Vit}. Note that in this representation, the state behind the
front at $\xi=- \infty$ corresponds to a point on the $\sigma $ axis,
and that the front solution $\sigma(\xi)$, $E(\xi)$ is represented in
this diagram by the flow along one of the trajectories towards either
the positive $E$-axis for $PSF$ or the negative $E$-axis for $NSF$ for
$\xi \rightarrow \infty$. Note
furthermore that $\sigma$ overshoots the value $\sigma^-$
($=\sigma(\xi \rightarrow -\infty$)) in the case of $NSF$. This
property as well as the monotonicity of $\sigma[E]$ and accordingly of
$\sigma(\xi)$ for positive fronts, follows immediately from Eq.\ 
(\ref{406}). For $NSF$, it can also be observed in the three-dimensional
simulations of Vitello {\em et al.} \cite{Vit}, shown in Fig.\ 1(c).

The smallest $E^+$ for which a front solution with $v=2$ is shown 
in Fig.\ 2(a), is $E^+ = -1.999$. For this value of $E^+$, 
$\sigma[E]$ continues to increase till $E \approx E^+$ and 
then suddenly decays to zero. A short analytical investigation 
of (\ref{406}) shows, that this behavior develops into a 
discontinuity of $\sigma[E]$ at the point $E=E^+$ for $v = -E^+$. 
$\sigma[E]$ then increases monotonically up to
$f(E^+)$ for $E \downarrow E^+$ and then jumps to zero
discontinuously at $E^+$. This shock-like behavior stays unchanged
under a parameter change to $\sigma(\xi)$. It is further discussed
and motivated in Section V.

An immediate consequence of Eqs.~(\ref{406}) and (\ref{407})
for the electron and ion density is that the value $\sigma^-$ 
behind the front (where $E \rightarrow 0$) is a simple function 
of the value $E^+$ of the field ahead of the streamer profile:
\begin{equation}
\label{4010}
\sigma^-(E^+) \;=\; \rho_{|E^+|}[0]\;=\; \int_0^{|E^+|}
dx\;\frac{f(|x|)}{x} \end{equation} 
The virtue of this expression for the electron density $\sigma^-$ 
far behind the front as well as of the expression (\ref{407}) 
for the ion density $\rho$ throughout the whole front, 
is that it is independent of the velocity $v$, hence 
independent of whichever front profile is selected, provided that 
the limit $D\rightarrow 0$ is smooth. We shall see later that 
this $D=0$ result remains relatively accurate for $NSF$ fronts 
with $D\lesssim 1$, and compare it to the results of the 
simulations \cite{DW,Vit} in Section VI. For $PSF$, on the other 
hand, the above result will turn out to be less relevant due
to the non-perturbative nature of the limit $D\rightarrow 0$ 
in this case.
 
For the Townsend function $f_T(|x|)$ [Eq.\ (\ref{2-14})] the function
$\sigma^-(E^+)$ can be expressed as \begin{eqnarray}
\label{4011}
\sigma^-(E^+)_T &\;=\;& |E^+| \;E_2\Big(|E^+|^{-1}\Big) 
\nonumber\\
&\;=\;& f_T(|E^+|)-E_1\Big(|E^+|^{-1}\Big)~, 
\end{eqnarray}
where $E_n(z)$ is the exponential integral \cite{Abramo}.

We finally note that the second form of (\ref{4011}) shows that
$\sigma^-$ approaches $f_T$ for large fields, since $f_T \gg E_1 $ for
$E^+\gg 1$. For $E^+$ of order unity, $\sigma$ and $f_T$ are still of
the same order, and this shows (for small $D$) that the growth rate
(\ref{2-17}) of long wavelength unstable modes in the unionized state
is comparable to the damping rate (\ref{2-19}) of stable modes in the
plasma behind a $NSF$. For small fields, the strict bounds
 on $E_2$ \cite{Abramo} show that $\sigma^- \approx E^+ f_T(E^+)$, so
that the approximate equivalence of these two time scales does not hold
for $E^+ \ll 1$, but in the small field range our starting model is
not very realistic anyway, because of the neglect of stabilizing
recombination terms.

\subsection{$D\neq0$ Front Solutions}

For $D\neq 0$, we can not obtain the uniformly translating solutions
analytically. Moreover, perturbation theory around the $D=0$ case is
not simply possible, as $D$ appears in front of the highest derivative
in Eq.~(\ref{3013}), so the diffusion term acts as a singular
perturbation. As a consequence, Eqs.\ (\ref{3013}) reduce to a set of
two coupled first order {\em ode's} for $D=0$, while three are
required for $D\neq 0$.  However, we can still easily demonstrate the
existence of a one-parameter family of uniformly translating front
solutions for $D\neq 0$ through standard counting arguments for {\em
  ode's}. Building on the results of such an analysis, the solutions
can then be constructed by integrating numerically in a stable
direction, using so-called ``shooting methods'' \cite{press}.

To perform the analysis, it is convenient to write the equations as a
set of three coupled first order {\em ode's}. There is some freedom
for the choice of the third variable: The standard choice would be
$\sigma' = \partial_{\xi} \sigma $, but for the discussion of the
singular limit as well as for numerical stability, the charge density
$q$ has turned out to be the most convenient choice.  The {\em
  ode's}~(\ref{3013}) then become \begin{eqnarray} \nonumber
\partial_\xi\; \sigma
&=& - \; {{\sigma E -v q}\over{D}}~,
\\
\label{503}
\partial_\xi \;E 
&=& \;\;\; q ~,
\\
\nonumber
\partial_\xi\;q 
&=& - \;\frac{\sigma f(|E|) }{v}\;+\; \frac {\sigma E -v q}{D}~.
\end{eqnarray} Just as we thought of the profiles for $D=0$ as 
describing flow in
a two-dimensional $(\sigma,E)$ phase space, we can now think of
Eqs.~(\ref{503}) as describing a flow in a three-dimensional ($\sigma,
E,q$) phase space. The velocity $v$ just plays the role of a parameter
in the flow equations, while $\xi$ again plays the role of a time-like
variable --- see the sketch in Fig.~2(b).

The steady states of the full {\em pde's} discussed in Section III
correspond to fixed points of the flow: the points $(\sigma,0,0)$ on
the $\sigma$-axis are fixed points of the flow (\ref{503}) and
correspond to homogeneously ionized plasma states, while the $E$-axis
is a line of fixed points ($0,E,0$) each of which corresponds to a
non-ionized state with $\sigma=\sigma' =0$ and $E\neq0$.

A uniformly translating front solution now corresponds to the
existence of a trajectory in this phase space that starts at ``time''
$\xi=-\infty$ on the $\sigma$-axis and flows to the $E$-axis for $\xi
\rightarrow \infty$. The multiplicity of such solutions (i.e., whether
they exist as discrete sets, or, e.g., as a one- or two-parameter
family) can be determined as follows. If we linearize the flow near an
arbitrary point $(\sigma^-,0,0)$ on the $\sigma$-axis by writing
$(\sigma,0,0)$ $= (\sigma^-,0,0)$ $+ A \exp({-\Lambda^- \xi})$, we
find the eigenvalue equation \begin{equation}
\label{5010}
\Lambda^- \; \left( {\Lambda^-}^2 - \Lambda^-\;\frac{v}{D} -
\frac{\sigma^-}{D} \right) \;=\; 0 ~.\end{equation} 
The fact that there is a zero
eigenmode is a consequence of the fact that the $\sigma$-axis is a
line of fixed points. For the two nontrivial eigenvalues (which
correspond to the linearized modes (\ref{2-19}) about the ionized and
screened region by equating $i {\bf k}\cdot \hat{\bf x}=-\Lambda^-$
and $\omega^- = \Lambda^-v$) we have \begin{equation}
\label{5012}
\Lambda^-_\pm \;=\; \frac{v \pm
\sqrt{v^2+4D\sigma^-}}{2D}~.  \end{equation}
 The eigenvalue $\Lambda^-_+$ is
positive, and hence gives a decaying exponential; thus points along
the corresponding eigendirection flow into the $\sigma$-axis as $\xi$
increases. The eigenvalue $\Lambda^-_-$, on the other hand, is
negative and hence corresponds to an unstable eigendirection, with
flow away from the axis. This implies that at each point
$(\sigma^-,0,0)$ on the $\sigma$-axis, there is, for fixed $v$, a
unique eigendirection $(-\Lambda_-^-,1,\Lambda^-_-)E_1$ along which the flow
is away from the axis. This flow can be followed in two anti-parallel
directions, determined by the sign of $E_1$. The one flowing towards
positive values of $E$ is the beginning of a $PSF$ front profile, the
one flowing towards negative $E$ is the beginning of an $NSF$ profile.
From these eigendirections, one derives that for $PSF$ with field
perturbations $E_1>0$,
the electron density decreases close to the $\sigma$ axis, while for
$NSF$ it increases. Accordingly, before reaching $\sigma=0$ for $\xi
\rightarrow \infty$, a $NSF$ profile has at least one maximum of
$\sigma$, while a negative one can be (and is) monotonic. This
generalizes our result for $D=0$, and is consistent with the 
findings of Vitello {\em et al.} \cite{Vit} shown in Fig.\ 1(c).
The physical origin of the maximum of $\sigma$ in
the rear end of the $NSF$ profile is the screening of the field: Due
to the low ionization rate in an already fairly suppressed field the
ion density has already almost acquired its final value, so the
electron density has to overshoot its asymptotic value $\sigma^-$ so
as to make $\partial_\xi E <0$. The screening behind a $PSF$ happens 
by suppressing the electron density faster than the ion density for
increasing $\xi$, and so there $\sigma $ is monotonic.

Let us now investigate the stability of the flow near a point
$(0,E^+,0)$ on the $E$-axis. Upon linearizing the flow equations
(\ref{503}) and writing the $\xi$ dependence of the perturbations in
the form $\exp(-\Lambda^+ \xi)$, we find the eigenvalue equation 
\begin{equation}
\label{506}
\Lambda^+ \; \left( {\Lambda^+}^2 - \Lambda^+\;\frac{v+E^+}{D} +
\frac{f(|E^+|)}{D} \right) \;=\; 0 ~.  \end{equation} 
Again, there is a zero
eigenvalue due to the fact that the whole $E$-axis is a line of fixed
points. The two nontrivial eigenvalues are \begin{equation}
\label{509}
\Lambda^+_\pm \;=\; \frac{v+E^+ \pm
\sqrt{(v+E^+)^2-4Df(|E^+|)}}{2D}~.  \end{equation} 
These eigenvalues can be
related to (\ref{2-17}) in the same way as (\ref{5010}) could be
related to (\ref{2-19}). For $v+E^+ >0$, the real parts of these
eigenvalues are always positive, so that both eigendirections are
stable. In other words, for $E^+ > -v$, all points near the $E$-axis
flow towards this axis --- in slightly more technical terms: there is
a two-dimensional stable manifold flowing into each of these points on
the $E$-axis. For $E^+ < -v$, the flow is away from the $E$-axis, and
fronts with $v+E^+<0$ can not be constructed. This generalizes
(\ref{409}) to $D\neq 0$.

The existence of a one-parameter family of fronts with velocity $v>
-E^+$ can now simply be understood as follows. As we saw before, there
is one unique $PSF$ and one unique $NSF$ trajectory flowing out of
each point on the $\sigma$-axis for fixed $v$ and $D$. Since the flow
defined by Eqs~(\ref{503}) is continuous, we can expect each
trajectory to extend smoothly \cite{footnote6b}. Once the flow gets
near the $E$-axis, we know from the above analysis that the trajectory
will be attracted completely to the axis, provided $v$ is large
enough. Thus, for each $\sigma^-$ and $v$, there will exist two unique
trajectories, i.e., a unique $PSF$ solution and a unique $NSF$
solution. Since each of these trajectories flows into a unique point
on the $E$-axis, the flow equations implicitly define a unique
relation of the form $\sigma^-=\sigma^-(v,E^+)$ for each of the two
types of fronts. For a given value of $E^+$, we thus have a
one-parameter family of front solutions, parametrized by $v$.

There are two important properties of the front solutions associated
with their asymptotic large $\xi$ behavior. First of all, we note that
according to (\ref{509}) the eigenvalues $\Lambda^+_\pm$ are only real
for \begin{equation}
\label{508}
v \;\ge\; v^* \equiv - E^+ + 2 \;\sqrt{D\;f(|E^+|)}~.  \end{equation} 
This implies
that the corresponding front profiles can certainly not approach the
asymptotic state $\sigma=0$ ahead of the front in a monotonic way for
$v<v^*$: When the eigenvalues are complex, the front profiles have an
oscillatory tail of the form $\exp[-(\Re \Lambda^+)\xi] \cos[(\Im
\Lambda^+) \xi) $. Clearly, this violates the physical condition that
the electron density $\sigma$ should remain positive, so solutions
with $-E^+ < v < v^*$ are physically excluded: $v^*$ denotes, in the
present case, the smallest velocity of physically allowable uniformly
translating front solutions.

The identification of $v^*$ as a bound on the velocity of physically
allowed front profiles depends only on the structure of the
eigenvalues $\Lambda^+$ associated with the {\em linear} flow near
unstable states. There is a second, {\em nonlinear}, way in which the
range of physically allowed values of $v$ can be bounded. To
understand this, note that for any $v \geq v^*$, the asymptotic decay
of $\sigma(\xi)$ for $\xi \rightarrow \infty$ for a uniformly
translating profile will be
\begin{equation}
\label{508a}
\sigma(\xi) = A_- \;\mbox{\large e}^{\textstyle -\Lambda^+_- \xi} 
\; + \; A_+ \;\mbox{\large e}^{\textstyle -\Lambda^+_+ \xi}
+ \mbox{h.o.t.} \end{equation} 
with real coefficients $A_-$ and $A_+$. Here,
h.o.t.\ stands for higher powers of the two exponentials generated 
when expanding the equations to higher than linear order in the 
variables. Clearly, the smallest eigenvalue $\Lambda^+_-$
governs the asymptotic decay of the profile provided $A_- \neq 0$.
That $A_-$ will generically be nonzero for an arbitrary velocity $v$
follows again from the counting argument above for the flow in phase
space: Each $PSF$ and $NSF$ trajectory flowing out of a point on the
$\sigma$-axis is unique, and hence there is no freedom to impose an
additional condition $A_-=0$ close to the $E$-axis. Furthermore, the
coefficients $A_-$ and $A_+$ depend on the full global nonlinear
behavior of the flow, and hence they depend implicitly on $v$.

There might exist, however, particular velocities $v^{part} > v^*$,
for which \begin{equation}
\label{420a}
A_-(v^{part})=0~.\end{equation} 
For discussing these we invoke again a continuity
argument for the front properties as a function of $v$. We expect a
very slowly decaying, nearly homogeneous uniformly translating front
solution to have a non-negative density everywhere, and to have a very
large velocity, since the velocity of a profile is essentially
inversely proportional to its slope in the limit of small slopes. 
(So indeed the roots $\Lambda_-^-$ given by (\ref{5012}) and
$\Lambda^+_-$ given by (\ref{509}) vanish in the limit that $v$
becomes large.) So for large $v$ we expect to find physical
solutions. These are characterized by $A_->0$ in the leading edge of
the front. Decreasing $v$ continuously, we either reach $v=v^*$
smoothly with still $A_->0$, or we reach the first particular velocity,
$v_1^{part}$, where $A_-$ vanishes. In the latter case, we expect by
continuity $A_-(v)<0$ for $v<v^{part}_1$. This implies that then
$\sigma$ approaches zero from below, i.e., that the front solution is
unphysical. Below the next $v_2^{part}$, we expect the electron
density to develop two zero's and so forth. The largest $v^{part}$, if
it exists, thus plays the role of the {\em nonlinear front velocity}
$v^\dagger$ \cite{vs3}, 
\begin{equation}
\label{4052}
v^\dagger = \mbox{max}\; \{\; v^{part}\; |\; A_-(v^{part})=0 \; \} 
\end{equation}
for a given $E^+$. 
(Note that if $\Lambda^+_- < 0.5 \Lambda^+_+$, the higher order
terms in (\ref{508a}) of order $\exp( -2 \Lambda^+_- \xi)$ are 
actually larger than the second term $\exp(-\Lambda^+_+\xi)$. 
This does not change our argument, though, as the prefactor 
of this second order term will vanish if $A_-$ vanishes.)

At the velocity $v=v^\dagger$ or at any $v=v^{part}$, the flow in 
phase space approaches the $E$-axis along the eigenvector where the 
flow is most rapidly contracting. The trajectory corresponding to the
nonlinear front solution is therefore more appropriately referred to
as a strongly heteroclinic orbit, where heteroclinic indicates that
it is a trajectory from one fixed point to another one. The
existence and properties of strongly heteroclinic orbits have
recently been under active investigation \cite{footnote8}.

Such a velocity $v^\dagger$, if it exists, bounds
the continuum of velocities of physical uniformly translating
solutions from below, and thus replaces the earlier bound $v^*$ 
derived from linearizing the equations in the leading edge of 
the front.
 
\subsection{Nonlinear Front Solutions for PSF}

For $NSF$, the bounding velocity $v^*$ given by (\ref{508}) is always
positive. Moreover, by integrating the flow equations (\ref{503})
numerically and searching for particular solutions for which,
according to (\ref{420a}), $A_-(v^{part})=0$, we have convinced
ourselves that there are no such solutions for any $D\neq 0$ and
$E^+<0$. Hence, the smallest velocity of physical $NSF$ solutions is
always $v^*$, for any value of the parameters.

For $PSF$, on the other hand, the situation is very different, since
$v^*<0$ for $(E^+)^2 >4Df(|E^+|)$ --- for the Townsend function
(\ref{2-14}), this happens for $D\le 0.25E^+e^{1/E^+}$, hence for 
any $E^+$ for $D\le 0.68$. In particular for $PSF$ at small $D$ the 
question therefore arises whether there are nonlinear front solutions 
defined by (\ref{420a}) and (\ref{4052}) with $v^\dagger >0$. The 
results of a numerical search for such solutions are shown in Fig.\ 3, 
as a function of $D$ and $E^+$. Below the full line in this diagram, 
there exists indeed a nonlinear front $v^\dagger >v^*$, whereas above 
this line $v^*$ denotes the smallest velocity of physical front 
solutions. While these results have been obtained numerically, the 
existence of a single (unique) particular solution with 
$A_-(v^{part})=0$ in the limit $D\rightarrow 0$ can be demonstrated 
analytically. Since a full discussion of these results will be given 
elsewhere \cite{ebert}, we confine ourselves here to a brief outline 
of the arguments that also demonstrate the singular nature of these 
solutions for $D\rightarrow 0$.

If we take the limit $D\rightarrow 0$ with $v$ fixed, assuming no
nontrivial scaling of the variables $\sigma$, $E$ and $q$ and of the
spatial coordinate $\xi$, Eqs.\ (\ref{503}) can easily be shown to
reduce to those studied in Section IV.A for $D=0$. Hence, we can recover
in this way the family of front solutions obtained there. {\em Any}
particular solution, on the other hand, for which $A_-(v^{part})=0$,
decays according to (\ref{508a}) as $\exp(-\Lambda^+_+ \xi)$ as $\xi
\rightarrow \infty$. Since $\Lambda^+_+ \propto D^{-1}$ for
$D\rightarrow 0$, such a particular front solution becomes extremely
steep as $D\rightarrow 0$: its gradients diverge as $1/D$ and so that
the diffusion term can still overcome the drift term as $D\rightarrow
0$. That the velocity of such a solution must also have a nontrivial
scaling in this limit can be seen from the third equation of
(\ref{503}), written in the form
\begin{equation}
\label{4053}
\partial_\xi\;q 
= \sigma \; \left( -\;  \frac{ f(|E|) }{v} + \frac{E}{D} \right) \;-\;
\frac {v}{D}\; q~.  \end{equation} 
Any {\em nontrivial} scaling of this equation in the
limit $D\rightarrow 0$ can only occur if the first term between
brackets remains of the same order as the other two, which diverge as
$1/D$. This is only possible if $v$ scales as $D$. In this limit, the
third term can then be
neglected, and since $\partial_\xi q$ has to change sign in the front
region (as the charge density $q$ vanishes as $\xi \rightarrow \pm
\infty$), there must be an intermediate value $\hat{E} <E^+$ of the
field for which $v=Df(|\hat{E}|)/\hat{E}$.

Now that we know the scaling of the spatial gradient of the velocity
of such particular front profiles for $D\rightarrow 0$, one easily
convinces oneself that the electron and charge density of these
solutions must {\em diverge} as $1/D$ in this limit. To study the
existence of such possible solutions, it is therefore convenient to
introduce new variables and coordinates according to \begin{equation}
\label{4054}
x=D\tilde{x}~,\;\;  v=D\tilde{v}~,\;\; \xi=D\tilde{\xi}~,\;\;
\sigma=\tilde{\sigma}/D~,\;\;  q=\tilde{q}/D~, \end{equation} 
with $E$ and $\tau$ unchanged. In these new variables, the flow 
equations (\ref{503}) become 
\begin{eqnarray} \nonumber
\partial_{\tilde{\xi}}\; \tilde{\sigma}
&=& - \; \tilde{\sigma}\; E + D \; \tilde{v}\; \tilde{q},
\\
\label{4055}
\partial_{\tilde{\xi}} \; E 
&=&\tilde{ q} ~,
\\
\nonumber
\partial_{\tilde{\xi}}\;\tilde{q} 
&=& \;\tilde{\sigma} \left( E- \frac{f(|E|)}{\tilde{v}}\right) \; - \;
D \; \tilde{v}\; \tilde{ q}~.  
\end{eqnarray} The limit $D\rightarrow 0$ can now
be taken simply by leaving out the term $D\tilde{v}\tilde{q}$ in the
first and last equation. We will show elsewhere \cite{ebert} that the
resulting equations have {\em one unique} physical front solution thus
fixing one particular value of the scaled velocity $\tilde{v}_1$ and
in view of the scaling (\ref{4054}) and the scaling of the eigenvalues
$\Lambda^+_{\pm}$, this solution must have $A_-(\tilde{v}_1)=0$. This
solution is therefore precisely the $D \rightarrow0$ limit of the
nonlinear front solution with velocity $v^\dagger =\tilde{v}_1 D$.
Furthermore, since the limit $D\rightarrow 0$ is smooth for Eqs.\ 
(\ref{4055}), this shows that there exists a nonlinear front solution
with $v^\dagger>0$ for {\em any} $E^+$ and nonzero but small $D$. Due
to the singular scaling (\ref{4054}), the corresponding front
solutions are determined by {\em ode's} that have a different
structure from those studied for $D=0$ in Section IV.A, and therefore
these nonlinear front solutions can not be obtained perturbatively
from the latter class of solutions --- of course, the latter class of
solutions still exists for $D\neq 0$, in agreement with the counting
arguments given earlier, but these now correspond to a singular limit
of Eqs.\ (\ref{4055})! The significance of these nonlinear front
solutions lies in the fact that they will turn out to be
the selected fronts that dominate the dynamics of $PSF$ in the
physically important range $0.1 \lesssim D \lesssim 0.3$.

The nonlinear front solution can be constructed numerically very
easily by integrating Eqs.\ (\ref{4055}) using standard numerical
``double shooting'' routines \cite{press}. Fig.\ 4 shows our numerical
results for the smallest physical velocity, max($v^\dagger,v^*$) in
the case that the ionization function is given by the Townsend
expression. The scaled velocities $v^\dagger/D$ and $v^*/D$ are
plotted; in agreement with our arguments above the scaled velocity
$v^\dagger/D$ of the nonlinear front solution approaches a finite
limit as $D\rightarrow 0$. Furthermore, the ratio $v^\dagger/D$ hardly
varies with $D$ in the physical range $0.1 \lesssim D \lesssim 0.3$,
and for small fields $E^+$, the scaled velocity $v^\dagger/D$ becomes
exponentially small, in agreement with the bound $v^\dagger/D < E^+
\exp(-1/E^+)$ that follows from the observations discussed after Eq.\ 
(\ref{4053}) above.

We finally note that our numerical routines have not only allowed us
to obtain the results show in Figs.\ 3 and 4, but have also enabled us
to verify numerically all the statements made above about the
multiplicity of solutions, the parameter ranges for physical fronts,
the monotony properties, the singular behavior of the small $D$ $PSF$
limit, and the persistence of the family of front solutions for
$D\rightarrow 0$.

\section{Selection of the Asymptotic Front}

\subsection{Front Propagation into Unstable States}

We have seen that the non-ionized state into which the streamer fronts
propagate, is an unstable state, that the homogeneous weakly ionized
plasma is a stable state, and that there is a family of uniformly
translating front solutions connecting the two. The existence of a
family of front solutions is a generic feature of front propagation
into unstable states. We, therefore, briefly recall what is known in
the literature for analogous problems and then translate this
experience to the streamer problem. The prototype equation for
studies of this type of front propagation is
\begin{equation}
\label{aw}
\label{3016}
\partial_t u = \partial_x^2 u + g(u)~,
\end{equation}
where $g(u)$ is some nonlinear function which satisfies
\begin{equation}
\label{gcond}
\label{3017}
g(0)=0~, ~~~g(1)=0~,~~~ g'(0) >0 ~,~~~
g'(1)<0~.
\end{equation}
Note that these relations imply that the ``state'' $u=0$ is unstable,
and that the ``state'' $u=1$ is stable. The study of the propagation
of fronts into the unstable state $u=0$ in this equation dates back to
the early work of Kolmogorov {\em et al.} \cite{kpp} and Fisher
\cite{fisher} in the context of population dynamics. Later Gel'fand
\cite{comb1} studied a particular example of a function $g(u)$
motivated by combustion. The mathematical research on this equation
culminated in the work by Aronson and Weinberger \cite{aw}, who
rigorously solved the front propagation problem for (\ref{aw}). In
particular, they proved that any initial perturbation, that is nonzero 
only in a finite part of space, approaches a unique uniformly 
translating front solution with velocity $v_f$ in the long time limit. 
If $g''(u)<0$ for all $u$, $v_f$ equals $v^*=2\sqrt{g'(0)}$ (derived 
from linearizing in the tip of the front), while for general $g(u)$, 
$v_f$ approaches either $v^*$ or some $v^\dagger>v^*$.  We refer to 
the literature for a detailed discussion of this work 
\cite{aw,eckmann}.

The velocities $v^*$ and $v^\dagger$ of the above problem 
directly correspond to our $v^*$ (\ref{508}) 
and $v^\dagger$ (\ref{4052}), since they are also the smallest 
velocities, which still allow for uniformly translating fronts
with $u\ge 0$ everywhere. So if $u$ is interpreted as a population
density or a chemical concentration, the selected front 
for every sufficiently localized initial state is the slowest
physical uniformly translating front. In other interpretations
no physical constraints bind $u$ to positive values.
Nevertheless the selected velocity stays the same. In this case,
one can prove that every front with smaller velocity is dynamically
unstable \cite{benjacob}, i.e., that the selected front is
marginally stable. The slowest physical or stable solution,
which is selected, coincides with the steepest physical or stable
one.

In the last decade, it has been recognized that several aspects of the
front selection problem encountered for the nonlinear diffusion
equation (\ref{aw}), seem to have more general validity. Certain
scenarios, justified by heuristic arguments but lacking a detailed
mathematical proof, were formulated and numerically tested on more
complicated {\em pde}'s that were often of higher order in the spatial
derivatives \cite{deelanger,benjacob,vs1,vs2,vs3,Oono}.  Some of the
equations studied lead to non-uniformly translating fronts that leave a
nontrivial spatially periodic state behind
\cite{deelanger,benjacob,vs2,vs3,deevs}.  A particular scenario is the
one distinguishing between the so-called {\em linear marginal
  stability} regime where $v_f=v^*$ and the {\em nonlinear marginal
  stability} regime where $v_f=v^\dagger$ \cite{benjacob,vs1,vs2,vs3}.
These names stem from the fact that in this formulation, the two
regimes of front selection are related to the stability properties of
the front solutions --- in both cases, the selected front separates
stable front solutions from unstable ones. Applied to (\ref{3016}),
this scenario just provides an intuitive explanation of all the
well-known mathematical results.  For plasma physicists, it is worth
mentioning that dynamics in the linear marginal stability regime is
related to that determined by the ``pinch point analysis'' which was
developed in plasma physics in the late fifties \cite{bers,LLX,vs2}.

\subsection{Predictions for Streamer Fronts}

By extending the arguments in the appendix of \cite{vs2}, one may show
that in the streamer case just like in the case of the above problem
(\ref{3016}), all physical solutions, i.e., all solutions with $u\ge0$
resp.\ $\sigma\ge0$ everywhere, are stable. For a detailed discussion, 
we refer to \cite{ebert}. It can be argued \cite{vs1,vs2}, and proven 
for (\ref{aw}) \cite{benjacob}, that a sufficiently localized initial
condition will approach the physical uniformly translating front,
which is closest in ``phase space'', i.e., the steepest one.  Both
for (\ref{3016}) and for the streamer equations, the steepest 
uniformly translating physical front is uniquely defined. It is also 
the slowest one.

We can immediately prove this when initially $\sigma(x,\tau=0)=0$ for
$x>x_c$ for streamer fronts with $D=0$: In general, there is a front
solution for every $v\ge\max[0,-E^+]$, but now the only way in which
the electrons can enter the range $x>x_c$ is through electron drift
with velocity $-E^+$. Clearly, therefore, the asymptotic front speed
of a $NSF$ can only be $-E^+$, while a $PSF$ can not propagate at all.
If the initial electron density, however, decays exponentially, the
local electron density grows by drift and ionization, and the front
can move quicker than $-E^+>0$ for a $NSF$.

For $D\ne0$, we will here only conjecture the analogous statements,
and we will test them numerically in Section VI:
\begin{enumerate}
\item {\em Selected front velocity.} If the initial conditions are
  sufficiently localized, {\em the selected front is the slowest
    physically acceptable front solution}, i.e., the slowest front
  profile for which $\sigma(\xi)\ge0$ for all $\xi$. In view of the
  discussion of Section IV, this means that the selected front
  velocity $v_f$ is predicted to be \begin{equation}
\label{w1}
v_f= v^* = -E^+ +2 \sqrt{D f(|E^+|)} ~, \end{equation} 
except when there exists a
nonlinear front solution satisfying (\ref{4052}): In that case 
\begin{equation}
\label{w2}
v_f=v^\dagger ~.  \end{equation} 
Note that the result (\ref{w1}) ($v^*$ is
the {\em Linear Marginal Stability} value in the terminology of
\cite{vs1,vs2}) is an explicit expression for $v_f$ in terms of
parameters associated with the linear instability of the unstable
state only. On the other hand, the existence of a nonlinear front and
the value of $v^\dagger$ (the {\em Nonlinear Marginal Stability}
value) depends on the whole nonlinear behavior of the flow equations
(\ref{503}).
\item {\em Localized initial conditions.} Initial conditions are
  sufficiently localized if their spatial decay is faster than the
  asymptotic decay associated with the smallest eigenvalue
  of the selected profile, i.e., if 
\begin{eqnarray}
\label{w3} 
\sigma(x,\tau=0) &<& C \;\mbox{\large e}^{\textstyle 
-\Lambda^+_-(v^*) x} ~, \quad\mbox{or} \\ 
\label{w4}
\sigma(x,\tau=0) &<& C \;\mbox{\large e}^{\textstyle
-\Lambda^+_-(v^\dagger) x} ~, \\
&&\mbox{for } x\rightarrow \infty ~, \nonumber
\end{eqnarray} 
depending on whether the
selected front is $v^*$ or $v^\dagger$. Here $C$ is an arbitrary
constant, and $\Lambda^+_-(v^*)$ ($= \Lambda^+_+(v^*) $) and
$\Lambda^+_-(v^\dagger)$ are given by (\ref{509}).
\item {\em Non-localized initial conditions.} If an initial condition
  does not obey (\ref{w3}) or (\ref{w4}), faster front speeds are
  possible. In particular, if initially $\sigma(x,\tau=0) \sim
  \exp(-\Lambda x)$, with $\Lambda < \Lambda^+_-(v^*)$ or $\Lambda <
  \Lambda^+_-(v^\dagger)$, whichever regime applies, then the front
  speed is given by \begin{equation}
\label{w5}
v= -E^+ +D \Lambda + \frac{f(|E^+|)}{\Lambda} ~, \end{equation} 
which is obtained by solving (\ref{506}) for $v$ in terms of $\Lambda$.
\end{enumerate}

We now combine the analytic and numeric findings from Section IV with
the selection rules above to quantitative predictions for asymptotic
fronts, which evolve from sufficiently localized initial conditions,
in the case that the impact ionization is given by the Townsend
expression (\ref{2-14}):
\begin{itemize}
\item[$NSF$:] For $NSF$, we numerically have not found any nonlinear 
  fronts for any $D$ and $E^+$, so our simple yet 
  powerful prediction is that for $NSF$ $v_f=v^*$ with $v^*$ given by
  (\ref{w1}). In principle it is possible that for
  other ionization functions $f(E)$ than the Townsend function
  (\ref{2-14}), there can be nonlinear front
  solutions also in the $NSF$ regime. In practice, we expect, however,
  that this will not be the case for physically reasonable functions
  $f(E)$, i.e., for functions consistent with (\ref{2-17a}).

  Once the predicted velocities are known, the value
  $\sigma^-$ of the electron density behind the streamer head is
  obtained from the numerical integration of the flow equations.
  The results of these calculations are shown in Fig.~5$(a)$.
  Since for $NSF$, the limit $D\to0$ is smooth, also $\sigma^-$
  depends only weakly on $D$ for $D \lesssim 1$, so that the $D=0$
  prediction (\ref{4011}) is quite accurate for realistic values of 
  the diffusion coefficient.

  At the predicted values of the selected front velocity, the
  width of the front region can be obtained directly from our
  numerical solutions of the flow equations. We have somewhat
  arbitrarily defined the width $w$ as the distance between the points
  where $\sigma$ is 90\% and 10\% of the value $\sigma^-$. As Fig.~6
  shows for $NSF$ fronts with $D=0.1$, this front width is typically
  of order 3 for field values of order unity. This confirms again that
  in the small $D$ limit the impact ionization length $\alpha_0^{-1}$
  sets the inner scale of streamer fronts. Furthermore, we find that
  our numerical data are well fitted by the expression $w \approx 6/
  \Lambda^+_\pm(v^*)$, which shows that the front width simply scales
  with the spatial decay rate $\Lambda_+^+(v^*)=\Lambda^+_-(v^*)$ of
  the streamer profile in the leading edge.
  $NSF$ fronts always have a maximum of the electron density
  within the front. 

\item[$PSF$:] As we saw in Section IV, for $PSF$ with $D\lesssim 0.9$,
  there always is a nonlinear front solution with velocity $v^\dagger
  >v^*$. The prediction is that in this range the selected front
  solution is the nonlinear front solution, i.e.,
  $v_f=v^\dagger$. Values of $v^\dagger$ as as function of $D$ and for
  several values of $E^+$ were already given in Fig.\ 4. We also saw
  before that these nonlinear front solutions are singular in the
  limit $D\ll 1$, where $v^\dagger \approx D \tilde{v}^\dagger(D=0)$
  and $\sigma^{-} \approx \tilde{\sigma}^-(D=0)/D$. The resulting
  predictions for $\sigma^-$ are shown in Fig.\ 5$(b)$.

The fact that the dimensionless inner decay length of these nonlinear
fronts scales as $D$, implies that the physical decay length of such
solutions is $D/ \alpha_0= D_e/(\mu_e E_0)$, i.e., is given by the
electron diffusion length. However, since simultaneously the electron 
density $\sigma^-$ diverges as $1/D$, the total front width $w$ 
defined above still approaches a finite limit as $D\rightarrow 0$ 
in units of the ionization length $\alpha_0^{-1}$.

\end{itemize}

We finally note that the front propagation problem posed by the
one-dimensional streamer equations has a number of interesting
differences and similarities with the Aronson Weinberger front
propagation problem (\ref{aw}). In particular, it can be hoped, that
techniques of strict bounds developed for the time development of
these fronts \cite{aw} as well as for the nonlinear front velocity
$v^\dagger$ \cite{footnote8} might be also applicable to
planar streamer fronts.

\section{Numerical tests of the predictions}

We have tested the predictions listed in Section V by numerically
integrating the {\em pde's} (\ref{307}), (\ref{308}) forward in time.
Our computer program is a finite difference code with a time
integration which is based on a semi-implicit method.

We have performed an extensive search through parameter space, varying
$D$ between 0.02 and 3, and $|E^+|$ between 0.3 and 10. All our
numerical studies of the dynamics fully confirm our predictions for
fronts, and therefore we only present a sample of our results that
illustrate the important features.

All the simulations of the initial value problem we present in the 
remaining figures, have initially a field $E=-1$ constant in space. 
We keep the field constant in time in the non-ionized region. The 
simulations of Fig.\ 7 -- 10 start with the same localized initial 
ionization seed, a Gaussian profile for the electron density,
$\sigma(x,t=0) = 0.01 \exp -(x-x_0)^2$. Fig.\ 7 shows a run
for $D=1$ and times $t=0$ -- 130 in time steps $\Delta t = 2$. 
As can be seen, the small ionization seed near $x_0=50$ initially 
grows while drifting to the right in accord with Eq.\ (\ref{2034}). 
At time $t = {\cal O}(20)$, the ionization is strong enough that 
field saturation sets in and two asymmetrically propagating fronts 
emerge. The one propagating to the right develops into a uniformly 
propagating $NSF$ with velocity $v^*=2.21$ \cite{asymptotics} and 
degree of ionization behind the front $\sigma^-=0.130$. The maximum 
value of $\sigma$ in the rear part of the front is $\sigma_{max} = 
0.150$. At the same time, a structure develops on the left, which at 
time $t=130$ not yet has reached a stationary form, and which 
eventually will develop into a $PSF$. (Note that propagation to the 
left into a negative field $-E^+$ corresponds to a $PSF$ front moving 
to the right towards $x\rightarrow \infty$ in a field $+E^+$). How 
the $PSF$ actually reaches a uniformly translating profile, is shown 
in Fig.\ 8, where the development for $x_0=150$ and otherwise identical 
initial and boundary conditions is followed in time steps of $\Delta 
t = 10$ during the time $t=0$ -- 500. An asymptotic velocity of 
$v^\dagger = 0.22$ and a degree of ionization $\sigma^-=0.43$ is 
reached. Note the huge difference in the degree of ionization and  
in the front velocity already for the unrealistically large value 
of the diffusion constant $D=1$. 

The predictions from Section V for the selected uniformly 
translating fronts for $D=1$ and $E^+=\pm 1$ yield for the 
$NSF$ $v^* = 2.213$ and $\sigma^- = 0.129$, and for the $PSF$ 
$v^\dagger = 0.2199$ and $\sigma^- = 0.432$. They thus correctly 
predict the simulations of the initial value problem shown in 
Figs.\ 7 and 8 within the accuracy given. Note that for the velocity 
$v^\dagger$ of the $PSF$ and for the degrees of ionization $\sigma^-$ 
both behind the $PSF$ and the $NSF$ this fact also shows the relative
accuracy of the two very different numerical methods used, while 
for the velocity $v^*$ of the $NSF$ the numerical integration of the
initial value problem exactly reproduces the analytic result.

As $D$ decreases, both the structures within the fronts and the
asymmetry between $NSF$ and $PSF$ become more pronounced. 
We illustrate this in Figs.\ 9 and 10 with the temporal development
starting from the same initial perturbation as before, but now for 
$D=0.1$, the value corresponding to the parameter values of the
earlier three-dimensional simulations \cite{DW,Vit}. The time ranges 
in each plot are chosen appropriately for seeing the $NSF$ and the 
$PSF$ evolve into a uniformly translating state.
Fig.\ 9 shows a perturbation (initially localized at $x_0=50$)
evolving during time $t=0$ -- 130 in time steps $\Delta t = 2$.
Except for the smaller diffusion constant and the stretched
$x$ axis, the situation is thus identical with that of Fig.\ 7.
The $NSF$ on the right propagates with a somewhat smaller velocity 
$v^*=1.39$, leaving a slightly higher ionization $\sigma^-=0.147$
behind. The maximum $\sigma_{max}=0.199$ is relatively higher,
since diffusional smoothening of structures is less pronounced.
On the time scales of Fig.\ 9, the left front does not propagate,
but retracts into an apparently immobile structure. The electrons 
drift with the field into the ionized region, leaving a layer
of screening ions behind. Thus the electrons and the field are almost
separated such that ionization on this side almost cannot occur.
Eventually few electrons will reach the nonzero field region by
diffusion and slowly build up a higher ionization and ultimately
a propagating $PSF$. That a $PSF$ actually emerges, is shown in 
Fig.\ 10. Only times $t=4000$ - 8000 in time steps of $\Delta t =100$ 
after the initial perturbation at $t=0$ and $x_0=60$ are shown.
The front propagates with velocity $v^\dagger=0.0149$ leaving behind
an ionization $\sigma^-=6.32$.
The numerical values predicted in Section V are $v^*=1.384$ and
$\sigma^-=0.144$ for $NSF$, and $v^\dagger=0.0146$ and
$\sigma^-=6.234$ for $PSF$. The remaining numerical discrepancy of
maximally 2\% could be resolved by choosing a still smaller gridsize
in Figs. 9 and 10. Comparison of the $PSF$ for $D=1$ and $D=0.1$
indicates that the time it takes such a front to build up, rapidly
increases with decreasing $D$, but we have not pursued the scaling
of the transient time with $D$.

We finally show in Fig.\ 11 the evolution of streamer fronts starting
from non-localized initial conditions, i.e., not obeying the bounds
(\ref{w3}) or (\ref{w4}) for $D=0.1$. We used an initial electron 
density profile $\sigma(x,t=0)= 0.01/ (2\:\cosh\Lambda(x-200))$ with 
$\Lambda=0.25$ and an initial field $E=-1$. At these values, for the 
$NSF$, $\Lambda^{+}_{-}(v^*)=1.918$ and for the $PSF$, 
$\Lambda^{+}_{-}(v^\dagger)=0.3766$. 
In this case, the bounds (\ref{w3}) or (\ref{w4}) are indeed violated 
for both fronts, and Eq.\ (\ref{w5}) predicts a $PSF$ with velocity 
$v=0.497>v^\dagger=0.0146$ and an $NSF$ with velocity 
$v=2.497>v^*=1.384$. The simulations find the fronts propagating 
with velocities 0.50 and\ 2.50, respectively. The ionization behind 
the $NSF$ is $\sigma^-=0.149$ and behind the $PSF$ $\sigma^-=0.158$, 
so that now both are comparable to
each other and to $\sigma^-(D=0)=0.1485$ found from Fig.\ 5(a).
Note that the diffusion constant is identical with that of 
Figs.\ 9 and 10, the only difference being the extended initial 
perturbation.

The simulations confirm that streamer front propagation is indeed 
correctly described by the marginal stability scenario, which in the 
present case amounts to the statement that the slowest physical 
velocity is selected, whenever one starts from sufficiently localized 
initial conditions, just as for the simpler case (\ref{aw}).
 
\section{Conclusions and Outlook}

The analysis in this paper fully supports the validity of an
effective interface approach suggested by the results of the full
three-dimensional simulations of Dhali and Williams and of Vitello
{\em et al.} \cite{DW,Vit}. This emerges from our detailed study of
the associated one-dimensional problem, which yields the following
results:
\begin{itemize}
\item[{\em (a)}] After a very brief stage of transient exponential
  amplification of the initial ionized seed, the growing streamer
  evolves into an electrically screened plasma body separating two
  narrow fronts which propagate into the non-ionized outer region. We
  show that these two fronts correspond, for all practical purposes,
  to translating profiles which propagate independently. This entails
  that the separation of spatial scales between an inner front and an
  outer one, set by the global geometry, is indeed justified.

\item[{\em (b)}] This enables us to draw upon the existing knowledge
  about front propagation into unstable states and thus to provide
  definite predictions about:

$\bullet$ the relationship $v_f(E^+)$ between the velocity of a planar
  streamer front and the value of the electric field ahead of it, and

$\bullet$ the value of the degree of ionization of the plasma created
  by the front, $\sigma^-(E^+)$.

These predictions, although only valid as such in the absence of front
curvature, still compare very favorably with the numerical results of
Ref.\ \cite{Vit}. The two values of $\sigma^- (E^+)$ on the axis of Fig.\
1$(a)$ and 1$(b)$ behind the curved fronts of the 3D simulations
\cite{Vit} (with the convention that $E^+$ should be understood as 
the electric field value extrapolated from the external non-ionized 
region to the front position) are plotted in Fig.\ 5$(a)$. Without 
adjustable parameters our one-dimensional predictions for 
$\sigma^-(E^+)$ are well within a factor of 2 from the 3D simulations. 
Likewise, the velocity values for $v_f(E^+)$ even agree to about 20\%.
\end{itemize}

Moreover, our analysis shows that $NSF$ and $PSF$ propagate in 
this model and for realistic values of the reduced diffusion 
coefficient $D$, in a very asymmetric manner: 

$\bullet$ $NSF$ rapidly reach a regime of uniform propagation --- 
typically, on the scale of several tens of time units, i.e., in less 
than $10^{-10}$s. Their velocity is slightly larger than the electron 
drift velocity in the field $E^+$.

$\bullet$ This is to be contrasted with the dynamics of $PSF$: 
For realistic $D$-values, of order 0.1-0.3, they approach uniform 
translation considerably more slowly than $NSF$ --- typically on the 
time scale of $10^{-8}$s. Moreover, their asymptotic velocity is also 
much smaller than $v_f^{NSF}$. It obeys the inequality $v_f^{PSF}<DE^+
\exp(-1/E^+)$ \cite{ebert}. Finally, while the widths of $PSF$ and 
$NSF$ are comparable, the degree of ionization behind $PSF$ is much 
larger (up to a hundred times for $D=0.1$) than that behind $NSF$.

These results answer the question of whether $PSF$ do or do not
propagate, while explaining why the simulations of Vitello {\em et
  al.} \cite{Vit} could not yield  a definite answer --- most
probably, because, although their total width is of order
$\alpha_0^{-1}$, their true inner length scale, as defined by the
steepness of the profile, was too small to be resolved by their grid
size. (Note that the apparent symmetry between $PSF$ and $NSF$
found in earlier simulations \cite{DW} is to be related to the fact 
that there propagation into a pre-ionized medium (with initial 
electron density of $10^8/ \mbox{cm}^3$) is studied, and possibly 
also due to the use of a poorly resolving grid.)

It was observed empirically by Dhali and Williams \cite{DW}
that in the three-dimensional simulations, the dielectric relaxation 
time in the plasma behind the front was of the same order as the 
intrinsic time scale set by the front motion. Our analysis shows,
that this was no accident: It is a manifestation of the fact that
the balance of the growth mechanism (impact ionization) and the
stabilization mechanism (screening) leads to a single time scale
$t_0=(\alpha_0 \mu_e E_0)^{-1}$ for a $NSF$ and for
the relaxation behind it for fields of order $E_0$. Since our 
dimensionless value of $\sigma^-$ is the inverse dielectric 
relaxation time, it is of order unity (or slightly smaller) for 
fields $E^+ \approx -1$.

Of course, the above results should only be considered as a first step
towards a realistic treatment of streamer propagation. They will have
to be developed and extended along two different directions:

{\em (i)} {\em Predictions of patterns within the present model
and comparison with the simulations:}
Within the frame of the present continuous and fully
deterministic model, we here have only considered the restricted case
of a one-dimensional geometry. This enabled us to demonstrate that the
concept of effective interfaces does apply to streamers. This approach
will now have to be extended to the description of curved fronts. 
As also discussed in \cite{short}, one
will then be equiped with a reduced formulation, valid on the outer
scale, which will permit us to study real three-dimensional streamers
as pattern-forming systems, as was done, e.g., for viscous fingers and
dendritic solidification fronts \cite{reviews}. This should provide a
direct approach to the question of dielectric patterns, alternative to
the phenomenological $DLA$-inspired dielectric breakdown models
\cite{pietronero}.  

{\em (ii)} {\em Possibly, extensions of the model will be necessary 
to predict real experiments:}
We have based our analysis on the minimal model as defined in
Section II. It contains several restricting simplifications. A first
step in the improvement of the model would be to include the field
dependence of the transport coefficients $D_e$ and $\mu_e$. It is
clear that this will not modify our qualitative analysis, as, e.g.,
the counting argument for the existence of front solutions in Section
IV depends only on the linearization about the stable and unstable
states. Moreover, the qualitative asymmetry between the $NSF$ and
$PSF$ will persist as these result from the asymmetry of the electron
drift. Quantitatively, the value of $v^*$, the selected value of
$NSF$, will simply be given by (\ref{w1}) with the transport
coefficient and ionization rate evaluated at the field value $E^+$.
The slow transient build-up and small speed of $PSF$ could be affected
quantitatively by ionic motion, but from this effect, we expect no 
major qualitative differences. \\
Finally, it should be kept in mind that our continuum equations are
only valid on length scales larger than the mean free path
$\ell_{mfp}$. On the other hand, we find for the strongest field values
appearing in the simulations (which are much larger than the values of
the field across the gap, due to the enhancement near the streamer
tip), that the front width decreases down to about $3
\alpha_0^{-1}\approx 3 \ell_{mfp}$ in the approximation
(\ref{mfp}). In such limits, nonlocality of the transport and
ionization effects begin to play a role. In addition, 
under these conditions, a typical volume of size $\ell_{mfp}^3$ 
contains only of the order of 1000 electrons for the parameter 
values (\ref{2026}) used in the simulations.  Fluctuations are then 
likely to become non-negligible. In principle, treating these effects 
calls for a full kinetic description. This is probably out of reach 
for the moment, but one might want to mimic the main features of 
these effects by introducing stochastic terms in the equations.
These also could mimic photo-ionization somewhat before the front
due to photons released in the impact ionization events, or the
natural homogeneous background ionization due to radioactivity and 
cosmic radiation. Investigation of their relevance for branching of 
dielectric breakdown patterns might help to understand the asymmetry
between the macroscopic patterns of discharges propagating into a 
positive or a negative field \cite{maan}.

In conclusion, our analysis opens the way to a microscopically based
interface approach to discharges that seems promising for building a
coherent framework for the analysis of breakdown patterns of various
degrees of complexity.

\acknowledgements

WvS gratefully acknowledges hospitality of the Universit\'e Paris 
VII, where this work was initiated. UE thanks F.\ 
D\"obele and A.\ Stampa for valuable discussions about streamer 
experiments and plasma physics. Her work was made possible by the 
Priority Program Non-Linear Systems of the Dutch Science Foundation 
NWO. We also gratefully acknowledge financial support by NWO and the 
Lorentz Fund for visits of UE and WvS to the Universit\'e Paris VII. 
Finally, we like to thank P.A.\ Vitello for making copies of figures 
from \cite{Vit} available, which appear here as Fig.~1.

\appendix
\section*{Differences and similarities between combustion and streamer fronts}

In the introduction, we draw on the similarity between the streamer
problem and other problems like combustion, chemical waves, thermal
plumes, phase field models, {\em etc.}, to motivate the development of
an effective interface approach. Of these problems, streamer
propagation is most closely analogous to combustion, in that the
strong nonlinearity of the reaction rates (the combustion rate and the
ionization rate) is an important factor in giving rise to front
development in flames and streamers, respectively. There are important
differences as well, however, and since several interface techniques
were originally developed in the context of
combustion\cite{wil,buc,pf1}, we highlight some of the
differences and similarities here:

{\em (a)} In combustion the reaction rate depends strongly on the
temperature, whose outer dynamics is governed by a diffusion equation
of the form $\partial_t T\!=\!\nabla^2 T$, while for streamers the
ionization rate depends strongly on the field $|{\bf E}|$, with ${\bf
  E}$ the {\em gradient} of the potential $\Phi$ that obeys the
Laplace equation $\nabla^2\Phi \approx 0$ in the outer region where
the total charge density vanishes. This field strength ${\bf E}$
varies strongly in the streamer front, since the increased screening
resulting from the rising electron density suppresses ${\bf E}$ ---
and hence the ionization rate --- to zero. In combustion, on the other
hand, the temperature hardly varies throughout the combustion zone.

{\em (b)} Combustion fronts are essentially fronts progating
  into a metastable state, because the front has to supply the 
  heat that increases the temperature and hence the reaction rate,
  while streamer front propagation is an example of front propagation 
  into unstable states, where the reaction starts for any nonvanishing
  electron density. 

{\em (c)} In a
  flame front typically the temperature remains high enough that all
  the reactions proceed to saturation: all the combustable material
  burns.  The temperature difference between the flame front and the
  background is then essentially determined by conservation
  (conversion) of energy.  In typical streamer fronts, on the other
  hand, the field ${\bf E}$ is suppressed long before saturation
  effects start to play a role, and hence the ionization level behind
  the front is set by the internal dynamics of the front rather than
  by conservation laws (i.e., the gas density). 

{\em (d)} The electron
  drift $-\mu_e{\bf E}$ has no clear analogue in combustion.

{\em (e)} Finally, the relevant asymptotic expansion for streamers
  is not quite like the ``activation energy asymptotics'' of
  combustion \cite{wil,buc}, since we consider here fields strengths
  that are comparable to the characteristic field scale $E_0$ of the
  ionization rate given in Eq.\ (\ref{2006}) before the front, 
  whereas in combustion
  activation energy asymptotics is often appropriate since the flame
  temperature remains much smaller than the chemical activation
  energy. For streamers, an analysis like activation energy
  asymptotics is appropriate in the limit of small fields 
  $|{\bf E}|\ll E_0$.
  Of course, in streamers the rapid variation of the field ${\bf E}$
  in the front region, and hence the rapid suppression of the
  ionization rate, looks, at first sight, similar to the suppression
  of the chemical reaction rate with decreasing temperature in flames.
  However, in flames this is due to the strongly nonlinear
  dependence of the reaction rate on temperature before the front, 
  (so that a slight
  suppression of the temperature reduces the reaction rate
  dramatically), while in streamers in large external fields of order
  $E_0$ this is due to the fact that the field itself is reduced
  significantly behind the streamer front, as a result of screening.

\setlength{\unitlength}{1cm}

\begin{figure}[h]
\setlength{\unitlength}{1cm}
\begin{picture}(8,12.7)
\epsfxsize=6cm
\put(-.5,5){\epsffile{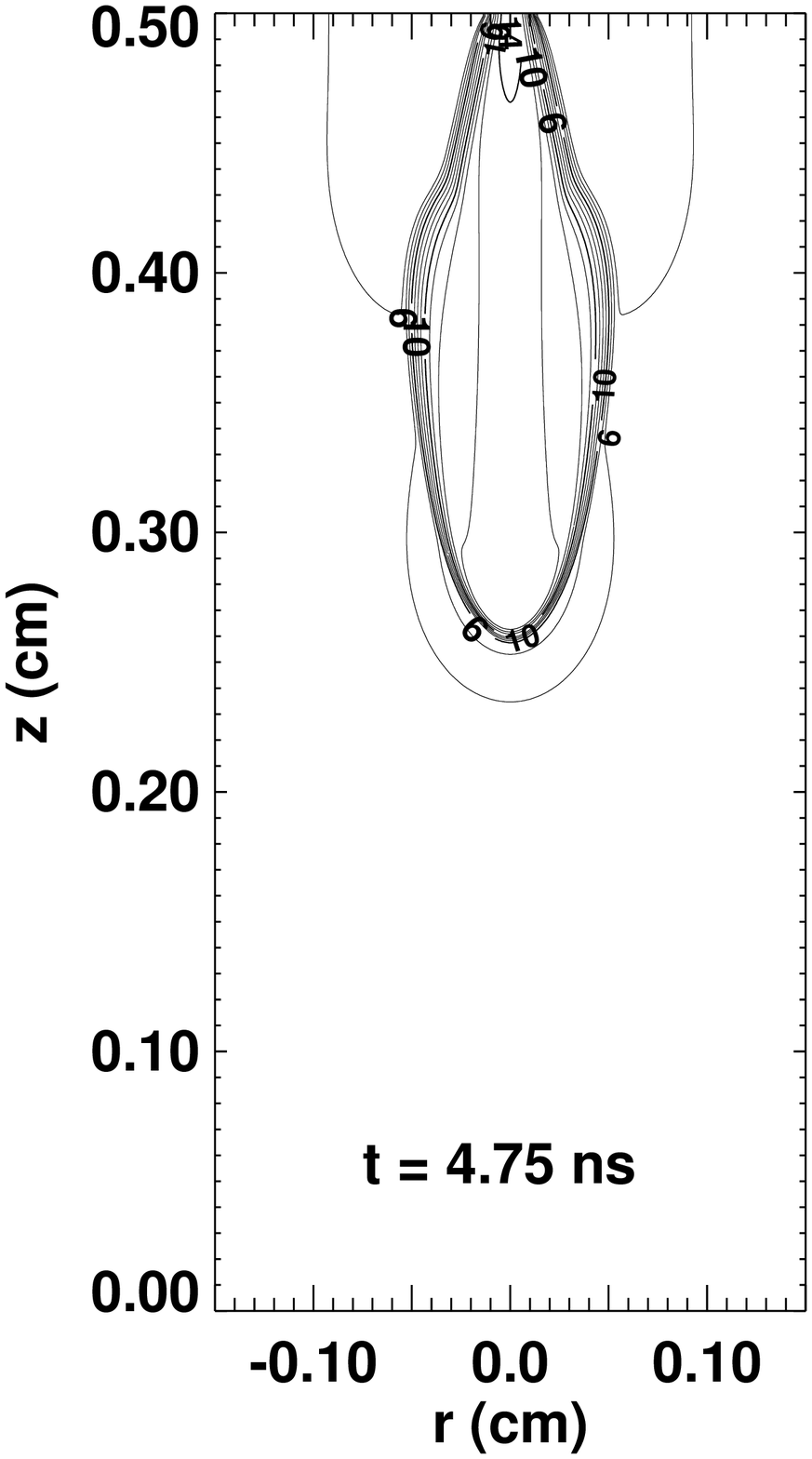}}
\epsfxsize=6cm
\put(3.5,5){\epsffile{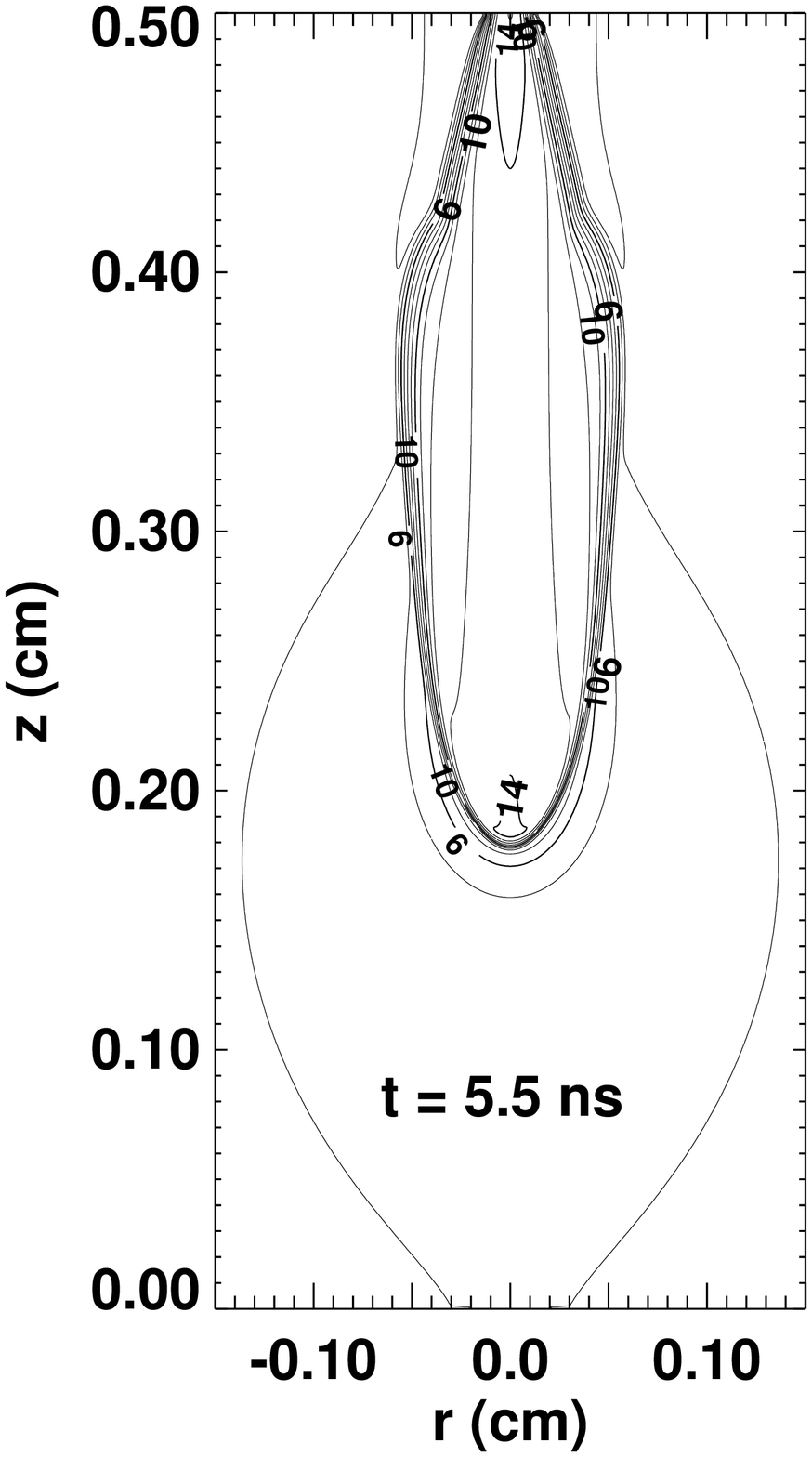}}
\epsfxsize=6cm
\put(-17,0){\epsffile{f1c.eps}}
\put(1.2,6.6){\large $(a)$}
\put(5.2,6.6){\large $(b)$}
\put(2.0,1.3){\large $(c)$}
\end{picture}
\begin{center}
\begin{minipage}{8cm}
\small FIG.\ 1. 
  Results of the numerical simulations of the full three-dimensional
  streamer equations (\ref{2011})-(\ref{2006}) of Vitello {\em et
    al.}, reprinted from Fig.~1 and 10 in \cite{Vit}.  $(a)$ Negative
  streamer propagating downwards towards the anode.  Electrodes are
  planar and located at z$=0$ and 0.5 cm; the voltage between the
  electrodes is 25 kV, which in the absence of the streamer amounts
  to a constant electric field $|{\cal E}|=E_0/4$. The system continues
  sidewards sufficiently far to make the lateral
  boundaries irrelevant. The streamer is assumed to be cylindersymmetric.  
  The dimensionless diffusion constant is $D=0.1$. Each line
  indicates an increase of $n_e$ by a factor 10; densities of $10^3 -
  10^{14}$ cm$^{-3}$ can be seen (initial background ionization: $1$
  cm$^{-3}$). Shape at time 4.75 ns after an initial ionization seed
  was placed near the upper electrode.  $(b)$ Shape at time 5.5 ns.
  $(c)$ Logarithmic electron $n_e$ and total charge $n_s$ density
  along the symmetry axis of $(b)$. Solid line: $n_e$; dot-dashed:
  $|n_s|$ for $n_s>0$; dotted: $|n_s|$ for $n_s<0$.  Note the
  exponential increase of the densities on $\mu$m scale within the
  front as well as maximum of both densities in the rear part of the
  front.  Courtesy of P.A.~Vitello.
\end{minipage}
\end{center}
\end{figure}

\begin{figure}[h]
\setlength{\unitlength}{1cm}
\begin{picture}(8,5.5)
\epsfxsize=6cm
\put(-.2,-2){\epsffile{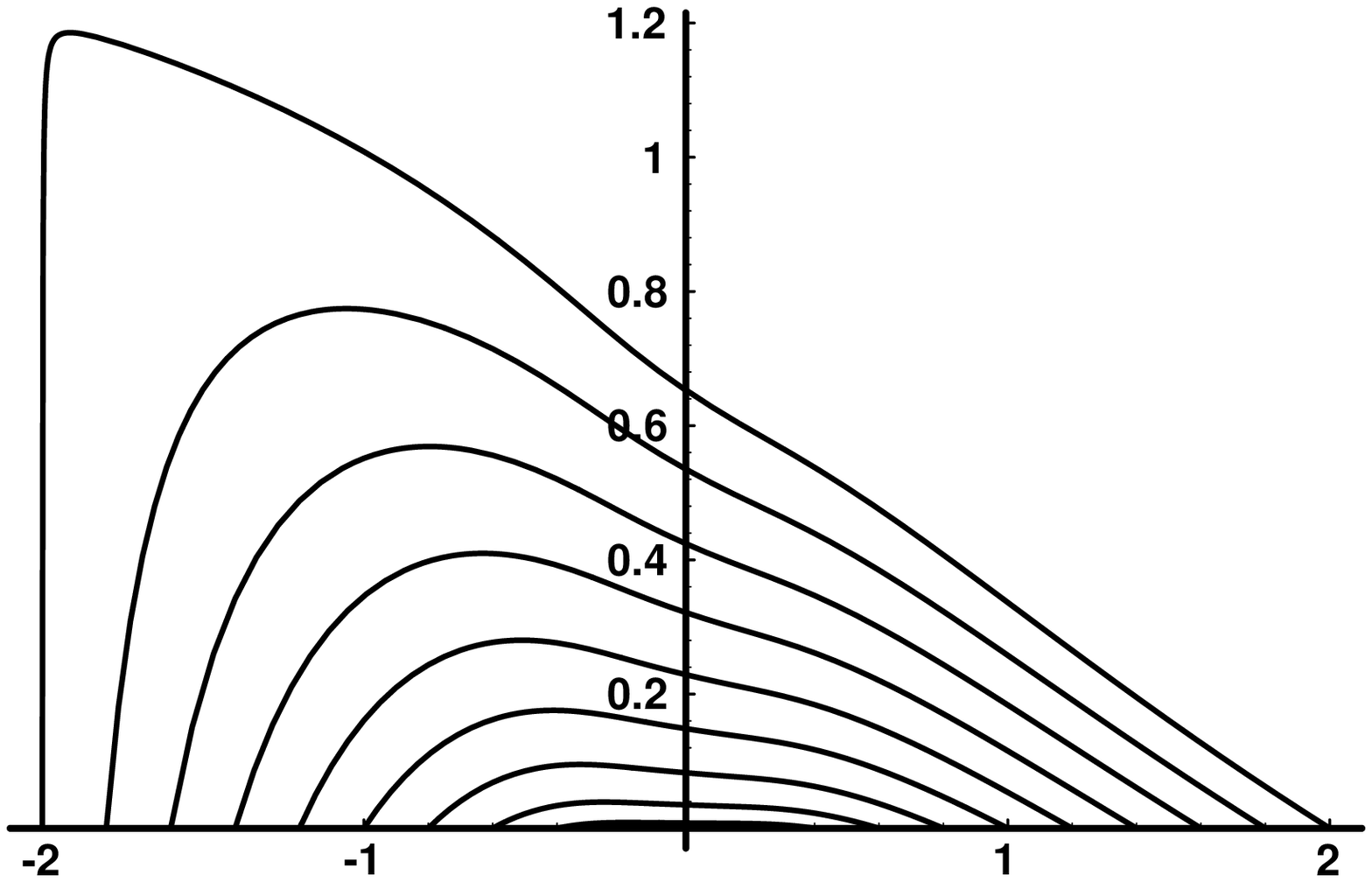}}
\epsfxsize=6cm
\put(3.0,-.5){\epsffile{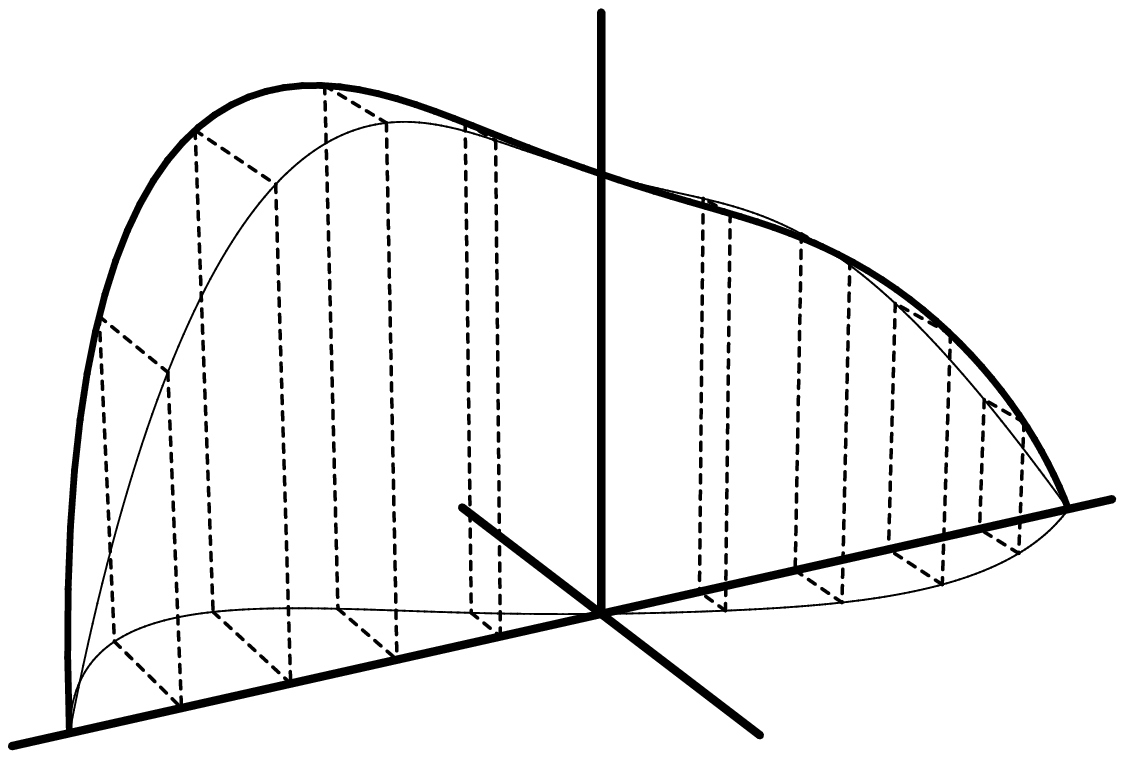}}
\put(.1,4.2){\large $(a)$}
\put(4.1,5){\large $(b)$}
\put(2.9,3.8){\large $\sigma$}
\put(5.65,0.5){\large $E$}
\put(6.1,4.8){\large $\sigma$}
\put(7.7,3.3){\large $E$}
\put(6.3,2.2){\large $q$}
\end{picture}
\begin{center}
\begin{minipage}{8cm}
\small FIG.\ 2. 
  $(a)$ Uniformly translating fronts for $D=0$ and $v=2$ shown as
  flows in the two-dimensional $(E,\sigma)$ phase space. Out of each
  point $\sigma^-$ on the $\sigma$ axis, there is a $PSF$ flowing to
  the right and a $NSF$ to the left. Both reach the same value $|E^+|$
  on the horizontal axis, which also is independent of $v$. Note, that
  $NSF$ have a maximum of $\sigma$ within the front, while $PSF$
  have monotonic $\sigma$.  Note also, that no physical fronts (i.e.,
  with $\sigma \ge 0$ everywhere) reach a value $E^+<-v=-2$, 
  in agreement with Eq.\ (\ref{409}).  
  $(b)$ Sketch of a uniformly translating $PSF$ and $NSF$ for $D \ne 0$
  as a flow in three-dimensional $(E,\sigma,q)$ phase space. The thick
  curves indicate the trajectories, while the thin ones show their
  projection into the $\sigma=0$ and $q=0$ planes. For fixed 
  $v$, there is at each point of the $\sigma$ axis still only one 
  outgoing vector, which can be followed in two antiparallel directions.
  The $E$ axis is fully attractive and will always be reached.
\end{minipage}
\end{center}
\end{figure}

\begin{figure}[h]
\setlength{\unitlength}{1cm}
\begin{picture}(8,5)
\epsfxsize=6.5cm
\put(.8,.3){\epsffile{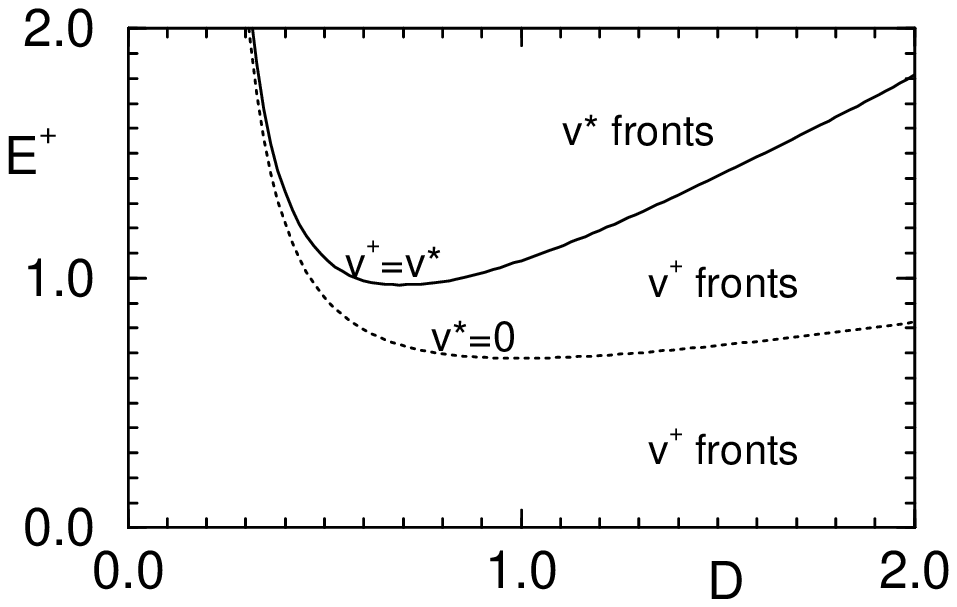}}
\end{picture}
\begin{center}
\begin{minipage}{8cm}
\small FIG.\ 3. 
  Phase diagram for $PSF$ as a function of $D$ and $E^+$. Above
  the solid line the lowest speed of physical front solutions is given
  by $v^*$, below the line $v^\dagger$ corresponds to the smallest
  speed of physical front solutions. Accordingly, the selected front
  speed is $v^*$ above the solid line ({\em linear marginal stability
    regime}), and $v^\dagger$ below the solid line ({\em nonlinear
    marginal stability regime}). The dotted curve indicates $v^*=0$
  and is a lower bound for the cross-over to $v^\dagger$ behaviour of
  the selected fronts.
\end{minipage}
\end{center}
\end{figure}

\begin{figure}[h]
\setlength{\unitlength}{1cm}
\begin{picture}(8,5.5)
\epsfxsize=7cm
\put(.5,.5){\epsffile{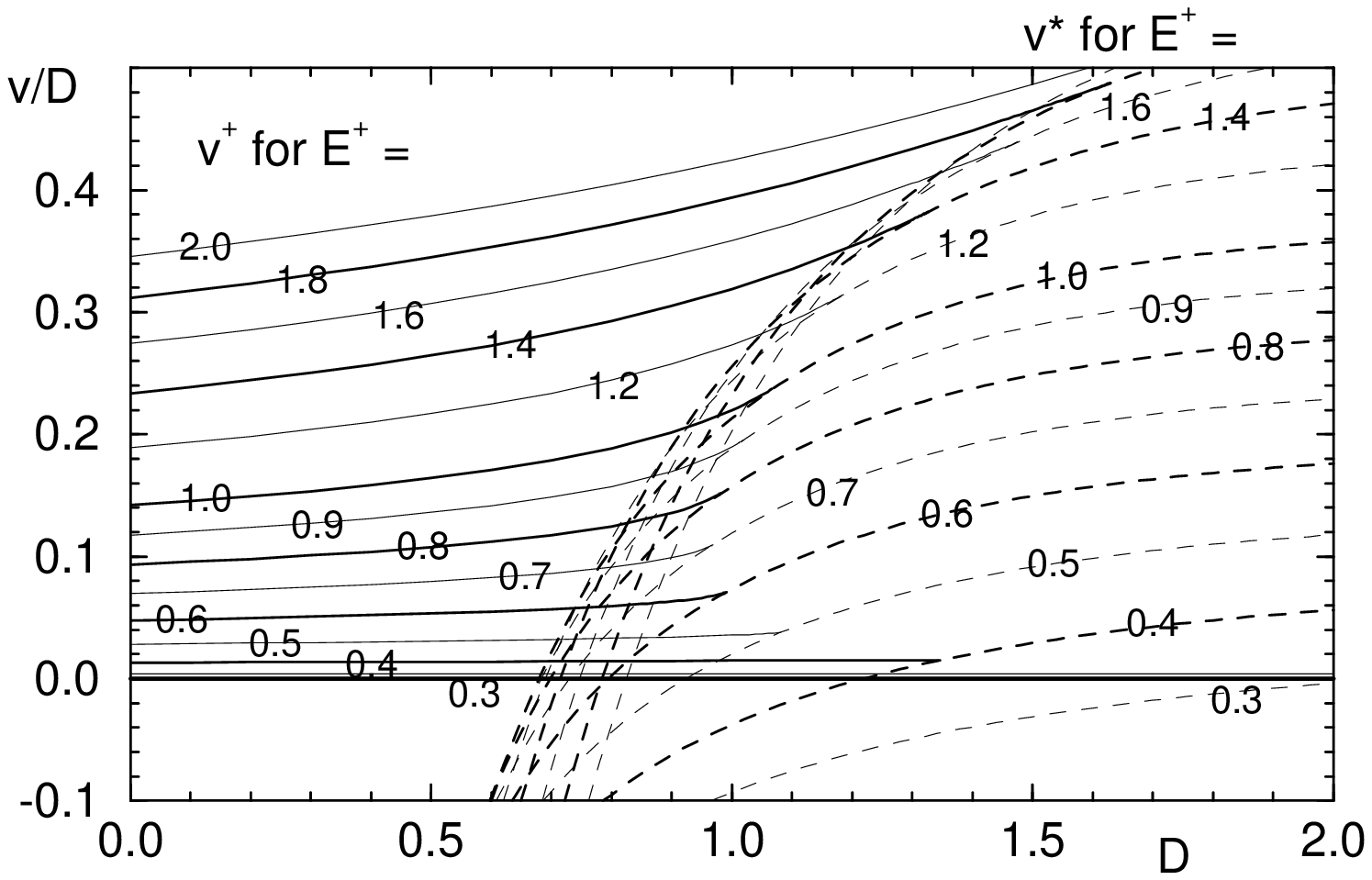}}
\end{picture}
\begin{center}
\begin{minipage}{8cm}
\small FIG.\ 4. 
  $\tilde{v}^\dagger = v^\dagger/D$ (solid) and $\tilde{v}^* = v^*/D$
  (dashed lines) as a function of $D$ for $E^+ =$ 0.3 -- 1.0 in steps 
  of 0.1, and for $E^+=$ 1.0 -- 2.0 in steps of 0.2. 
  $\tilde{v}^\dagger$ depends only weakly on $D$, i.e., the physical 
  front velocity $v^\dagger$ is approximately proportional to $D$. At
  $\tilde{v}^\dagger(E^+,D_{cr}(E^+))=\tilde{v}^*(E^+,D_{cr}(E^+))$, 
  the selected
  front crosses over from $v^\dagger$ to $v^*$; the $v^\dagger$ solutions
  disappear. Plotting $D_{cr}(E^+)$ in the $(E^+,D)$ plane yields the 
  solid curve in the phase diagram of Fig.~3, while the zeros of $v^*$
  determine the dotted curve in Fig.\ 3.
\end{minipage}
\end{center}
\end{figure}

\begin{figure}[h]
\setlength{\unitlength}{1cm}
\begin{picture}(8,10.5)
\epsfxsize=7cm
\put(.5,.5){\epsffile{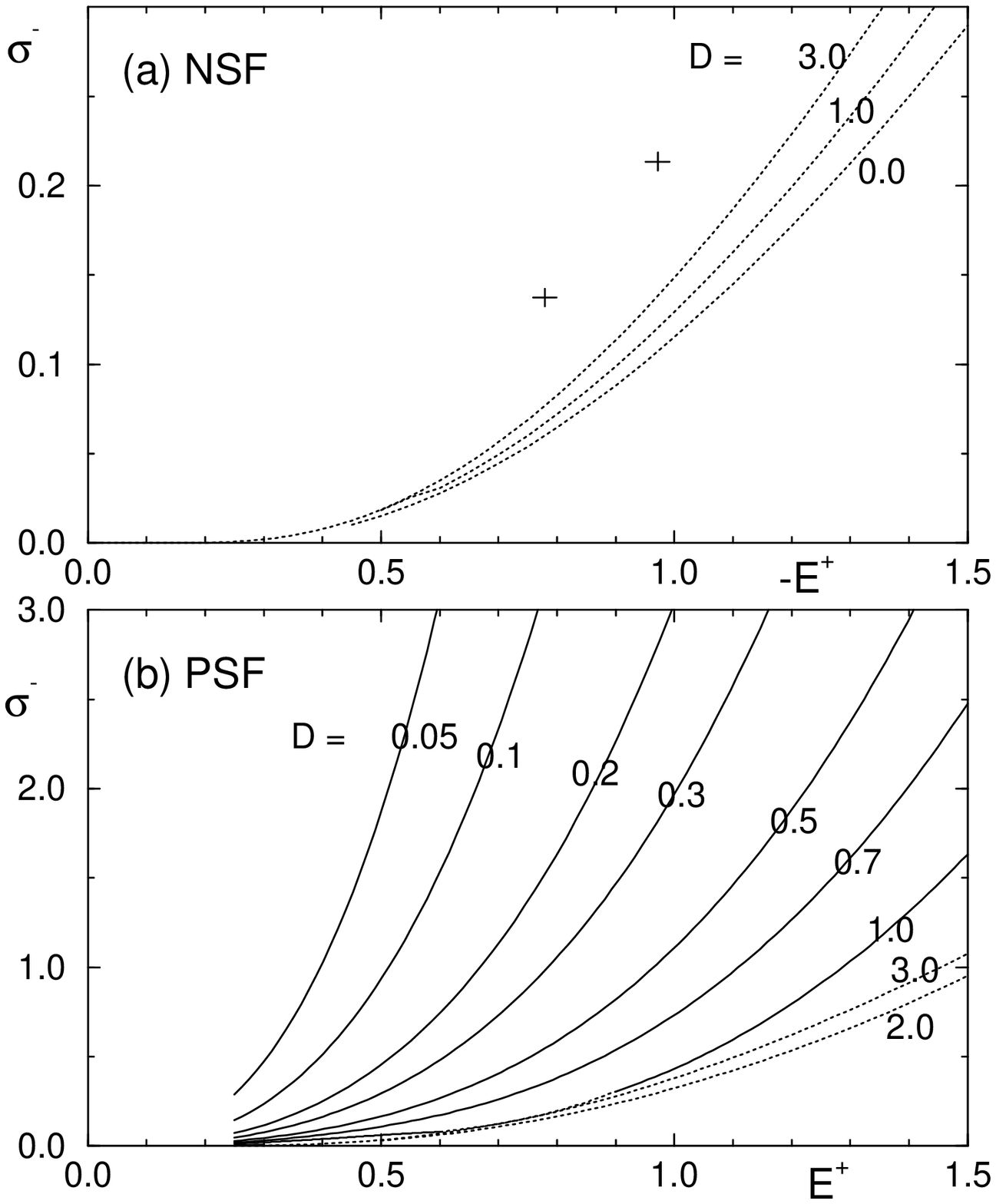}}
\end{picture}
\begin{center}
\begin{minipage}{8cm}
\small FIG.\ 5. 
  Electron density $\sigma^-$ behind the planar selected front as a 
  function of the field $E^+$ before the front for several $D$;
  dotted: $v^*$ fronts; solid: $v^\dagger$ fronts. 
  $(a)$ $NSF$: For $v^*$ fronts, $\sigma^-$ depends but weakly on $D$.  
  Results for $D=$ 0, 1, 3 are shown. Crosses: Extrapolation of 
  $\sigma^-(E^+)$  for $D=0.1$ for the curved fronts of the 3D 
  simulations \cite{Vit}. 
  $(b)$ $PSF$ results for $D=$ 0.05, 0.1, 0.2, 0.3, 0.5, 0.7, 1, 2, 3. 
  For $v^\dagger$ fronts, $\tilde{\sigma}^- = const. + {\cal O}(D)$,
  i.e., $\sigma^-(E^+)$ is approximately proportional to $1/D$.
\end{minipage}
\end{center}
\end{figure}

\begin{figure}[h]
\setlength{\unitlength}{1cm}
\begin{picture}(8,5)
\epsfxsize=7cm
\put(.5,.5){\epsffile{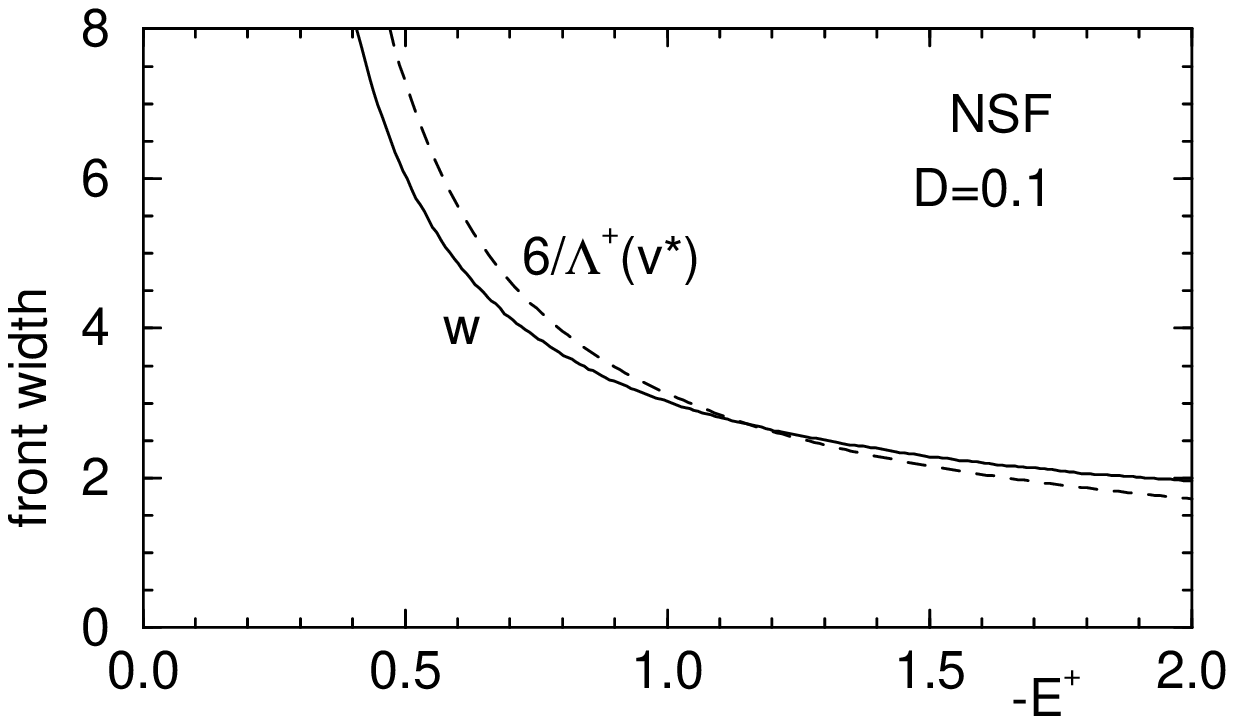}}
\end{picture}
\begin{center}
\begin{minipage}{8cm}
\small FIG.\ 6.
  Width $w$ of the front profile (measured between points
  with 0.1 $\sigma^-$ and 0.9 $\sigma^-$) as a function of $E^+$ for
  the selected $NSF$ fronts with $D=0.1$. The dashed line is given by
  $w=6/ \Lambda^+_-(v^*)$.
\end{minipage}
\end{center}
\end{figure}

\begin{figure}[h]
\setlength{\unitlength}{1cm}
\begin{picture}(8,9.5)
\epsfxsize=6.5cm
\put(.7,-.5){\epsffile{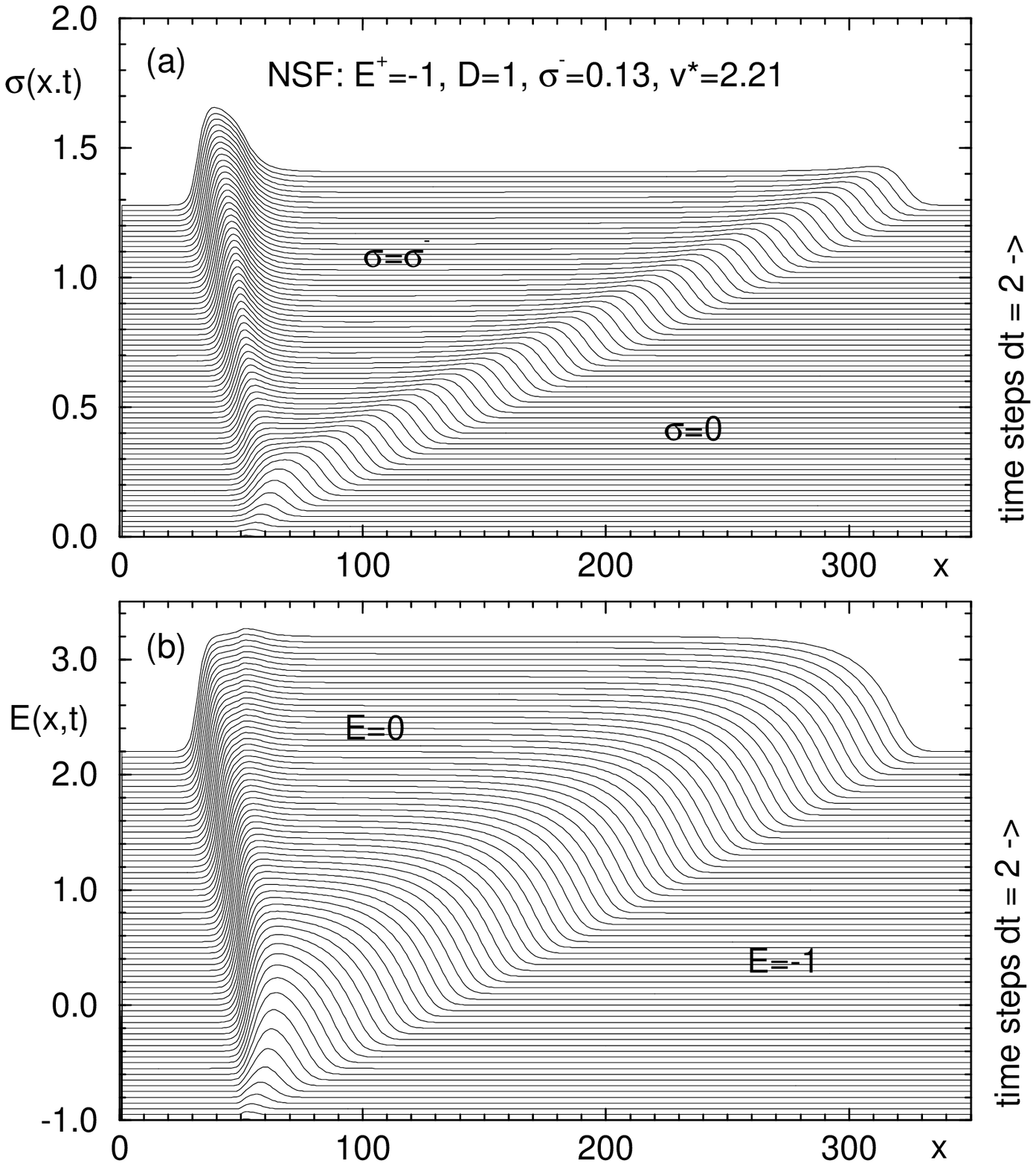}}
\end{picture}
\begin{center}
\begin{minipage}{8cm}
\small FIG.\ 7.
  Numerical integration of the time evolution given by
  Eqs.~(\ref{307}) and (\ref{308}) for $D=1.0$ in a constant
  background field $E=-1$ (numerical gridsize $\Delta x=0.1$ and
  timestep $\Delta \tau=0.05$, initial perturbation at $x_0=50$). 
  Initial condition at $t=0$: small charge-neutral, ionized region 
  of Gaussian shape depicted by the lowest line. Each new line 
  corresponds to a time step $\Delta t =2$ and the upper line to 
  $t=130$. $(a)$ The electron density $\sigma(x,t)$ initially grows
  and then, after field screening in the middle sets in, develops into
  a $NSF$ propagating to the right and a $PSF$ propagating to the
  left. $(b)$ The electric field $E(x,t)$ stays $E=-1$ in the 
  non-ionized region and becomes dynamically screened to zero in the 
  ionized region.
\end{minipage}
\end{center}
\end{figure}

\begin{figure}[h]
\setlength{\unitlength}{1cm}
\begin{picture}(8,5)
\epsfxsize=6.5cm
\put(.7,.5){\epsffile{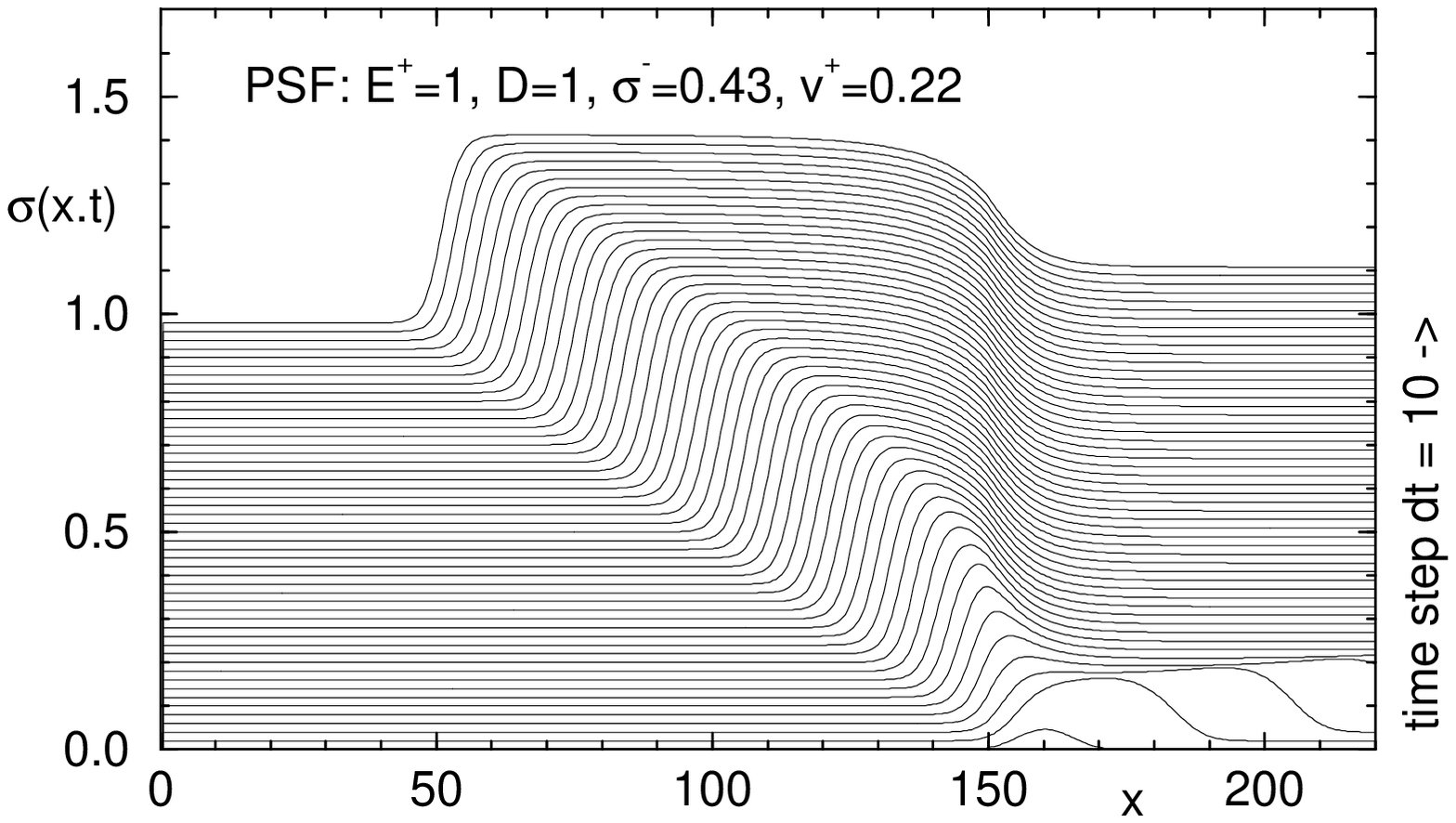}}
\end{picture}
\begin{center}
\begin{minipage}{8cm}
\small FIG.\ 8.
  Emergence of the uniformly translating $PSF$ on the left
  in the system of Fig.\ 7. Conditions identical to Fig.\ 7 except
  for $x_0=150$ and different numerical gridsize ($\Delta x=0.05$ and 
  $\Delta \tau= 0.01$). $\sigma(x,t)$ is shown in time range $t=0$ -- 500
  in time steps $\Delta t = 10$. Note the difference in the
  duration of the transient regimes, in the propagation velocities of
  $PSF$ and $NSF$, and in the degrees of ionization behind these.
\end{minipage}
\end{center}
\end{figure}

\begin{figure}[h]
\setlength{\unitlength}{1cm}
\begin{picture}(8,4.5)
\epsfxsize=8cm
\put(-.8,-1.5){\epsffile{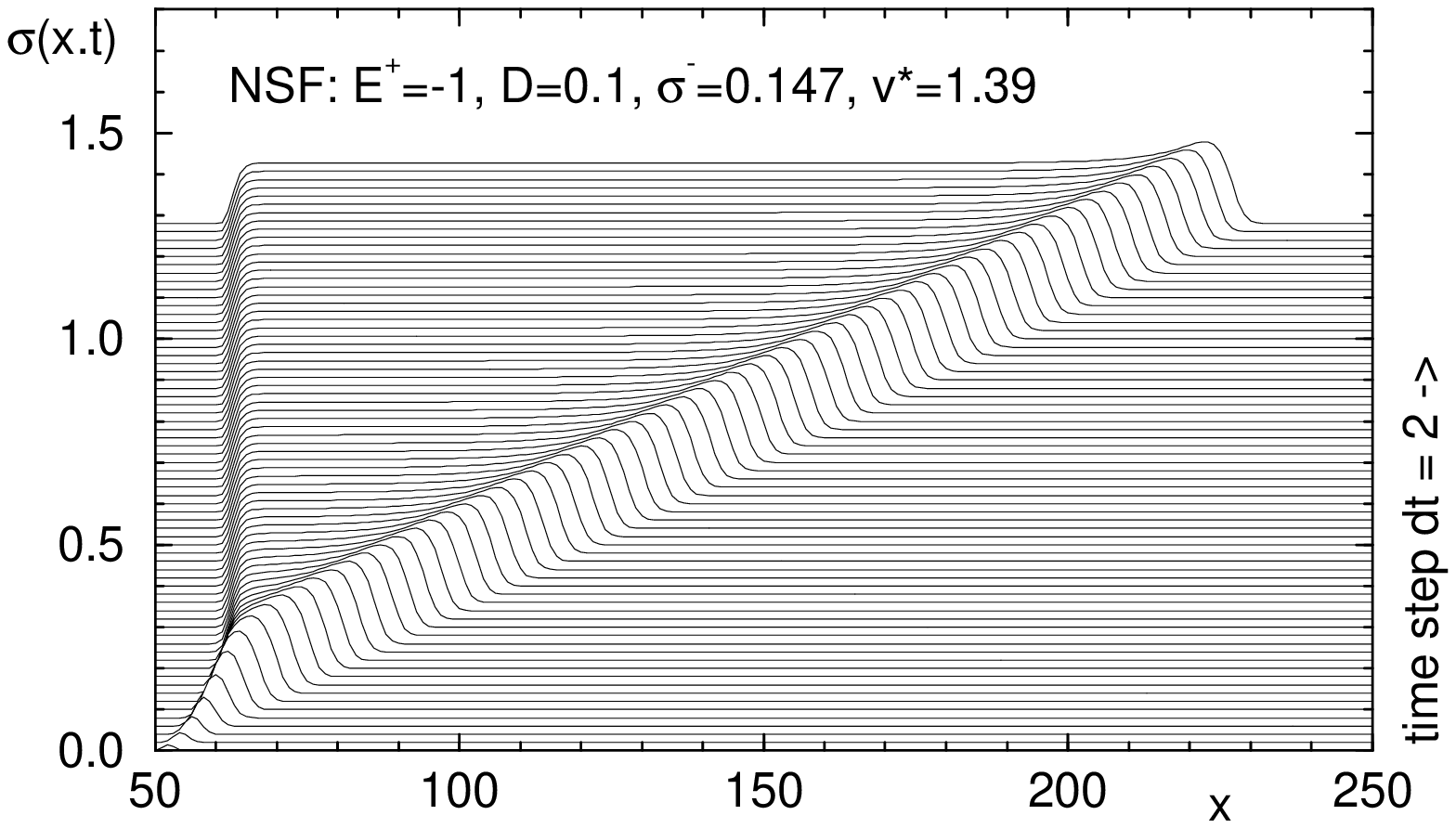}}
\end{picture}
\begin{center}
\begin{minipage}{8cm}
\small FIG.\ 9.
  Identical with Fig.\ 7$(a)$, except that here $D=0.1$. Time range also
  $t=0$ -- 130 in steps of $\Delta t = 2$. The $NSF$ has sharper contours 
  and propagates slower than for $D=1$, the $PSF$ appears not to develop.
\end{minipage}
\end{center}
\end{figure}

\begin{figure}[h]
\setlength{\unitlength}{1cm}
\begin{picture}(8,5)
\epsfxsize=6.5cm
\put(.7,.5){\epsffile{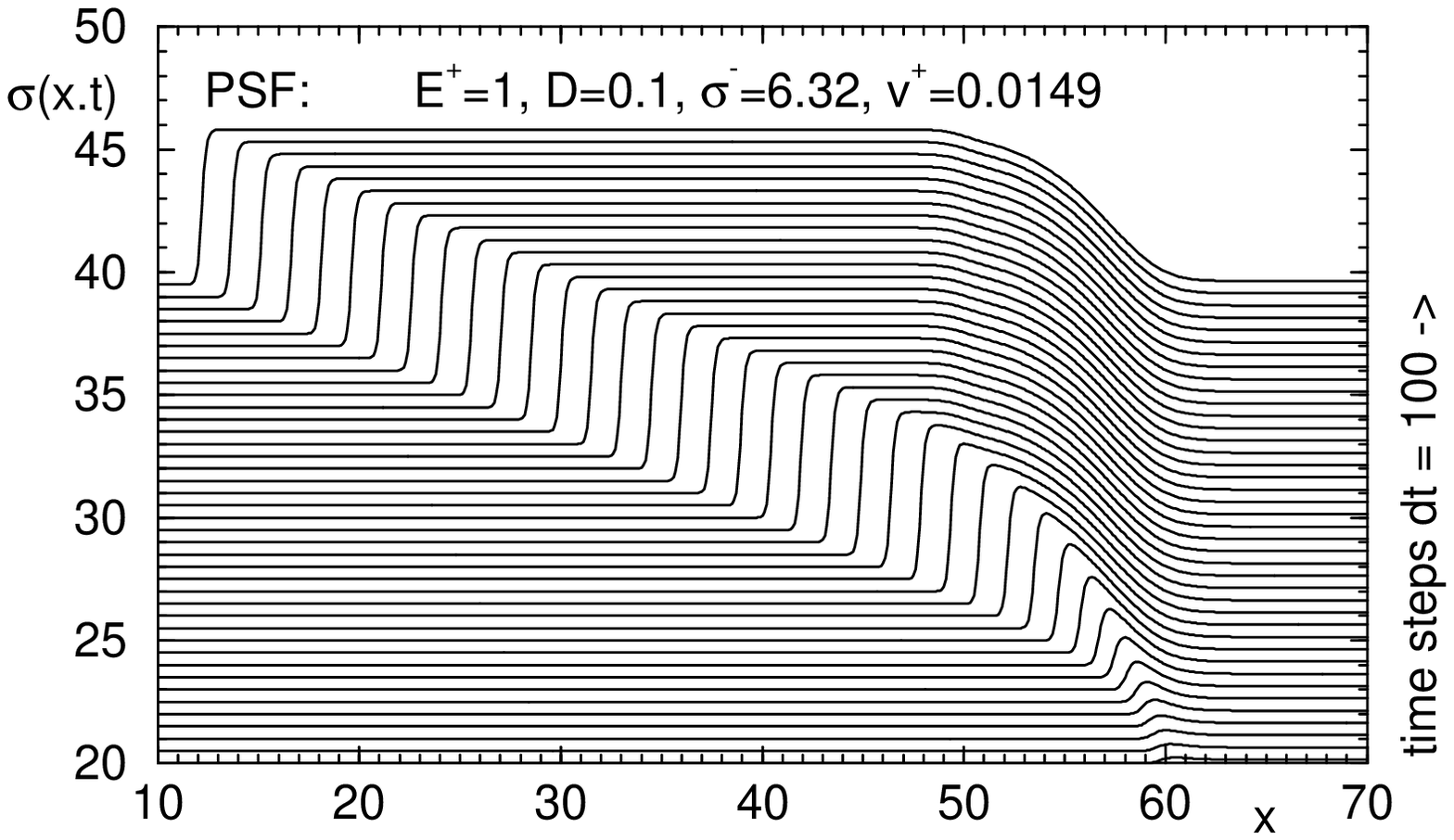}}
\end{picture}
\begin{center}
\begin{minipage}{8cm}
\small FIG.\ 10.
  Emergence of the uniformly translating $PSF$ on the left
  for $D=0.1$. Initial conditions identical with Fig.\ 9.
  The time range $t=4000$ -- 8000 after an initial
  perturbation at $t=0$ and $x_0=60$ is shown in time steps of 
  $\Delta t = 100$. (Numerical gridsize $\Delta x = 0.01$ and
  $\Delta \tau = 0.5$.)
\end{minipage}
\end{center}
\end{figure}

\begin{figure}[h]
\setlength{\unitlength}{1cm}
\begin{picture}(8,5)
\epsfxsize=6.5cm
\put(.7,.5){\epsffile{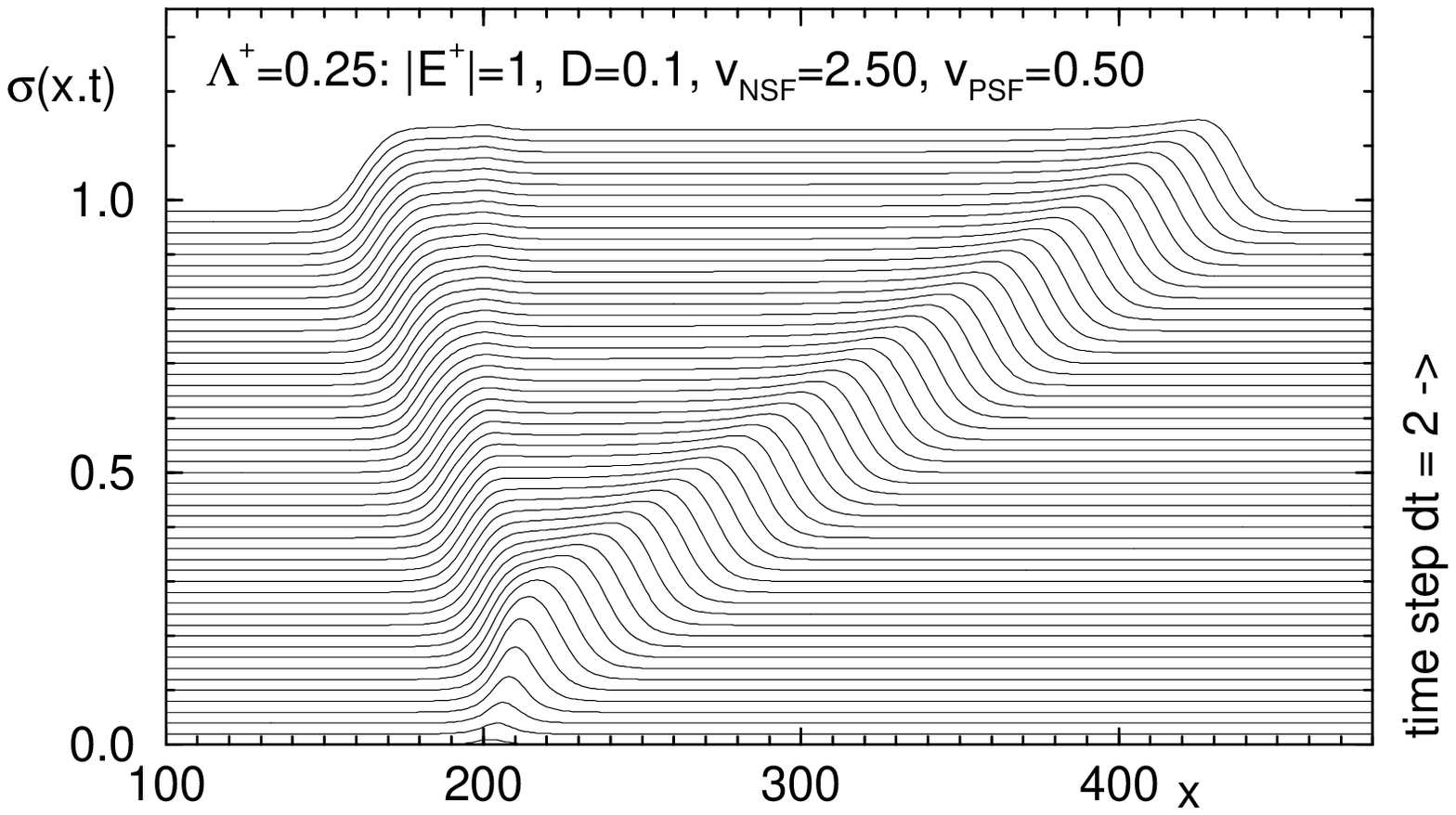}}
\end{picture}
\begin{center}
\begin{minipage}{8cm}
\small FIG.\ 11.
  A non-localized initial condition with $\Lambda = 0.25$
  as described in the text;
  otherwise, the situation is like in Fig.\ 9, and $D=0.1$.
\end{minipage}
\end{center}
\end{figure}
        
\end{multicols}


\begin{references}
\bibitem{footnote6b} Of course, in principle in a three-dimensional
  dynamical system, one can have limit cycles and other (strange)
  attractors. Limit cycles would in our case correspond to spatially
  periodic ionization waves, but the occurrence of such solutions is
  clearly excluded physically by the fact the ion density must be a
  monotonically increasing function of $\xi$. Other types of
  nontrivial behavior of the flow are excluded for the same reason.
\bibitem{asymptotics} By tracking the position $x_f(\tau)$ of the
  point where $\sigma(x_f(\tau),\tau)=0.1 \sigma^-$, we followed the
  convergence of the local velocity $v_f(\tau)=dx(\tau)/d\tau$ towards
  $v^*$, and found it to be consistent with the formula $v_f(\tau)=v^*
  -1.5D/(\sqrt{f|E^+|}\tau) + \cdots $ derived in \cite{vs2}.
\end{references}
\end{document}